\documentclass[11pt, oneside]{article}   	
\usepackage[margin=1in]{geometry}               
\usepackage{graphicx}				
										
\usepackage{amssymb}
\usepackage{physics}
\usepackage{amsmath}
\usepackage[dvips]{epsfig}

\usepackage{amsfonts}
\usepackage{subfig}

\usepackage{mathrsfs}

\usepackage{xcolor}

\usepackage{bm}

\usepackage{blindtext}

\usepackage[title]{appendix}

\usepackage{cite}

\usepackage{authblk}

\def\cchi{\raise2pt\hbox{$\chi$}} 

\title{\bf{Qubit Lattice Algorithm Simulations of Maxwell's Equations for Scattering from Anisotropic Dielectric Objects}}

\author {George Vahala ${ }^{1}$, Min Soe ${ }^{2}$, Linda Vahala ${ }^{3}$, Abhay K. Ram ${ }^{4}$, Efstratios Koukoutsis ${ }^{5}$, Kyriakos Hizanidis ${ }^{5}$\\
${ }^{1}$ Department of Physics, William \& Mary, Williamsburg, VA23185\\
${ }^{2}$ Department of Mathematics and Physical Sciences, Rogers State University, Claremore,OK 74017\\
${ }^{3}$ Department of Electrical \& Computer Engineering, Old Dominion University, Norfolk, VA 23529\\
${ }^{4}$ Plasma Science and Fusion Center, MIT, Cambridge, MA 02139\\
${ }^{5}$ School of Electrical and Computer Engineering, National Technical University of Athens,Zographou 15780, Greece \\}
\date{}

\begin{document}
\maketitle
\begin{abstract}
A Dyson map explicitly determines the appropriate basis of electromagnetic fields which yields a unitary representation of the Maxwell equations in
an inhomogeneous medium.
A qubit lattice algorithm (QLA) is then developed perturbatively to solve this representation of Maxwell equations.  
QLA consists of an interleaved unitary sequence of collision operators (that entangle on lattice-site qubits) and streaming operators (that move this entanglement throughout the lattice).
 External potential operators are introduced to handle gradients in the refractive indices, and these operators are typically non-unitary, but sparse matrices.  By also interleaving the external potential operators with the unitary collide-stream operators one achieves a QLA which conserves energy to high accuracy.  
 Some two dimensional simulations results are presented for the scattering of a one-dimensional (1D) pulse off a 
 localized anisotropic dielectric object.


\end{abstract}

\section{Introduction}

There is much interest in developing algorithms to solve specific classical problems that can be encoded onto a quantum computer.  One class of such algorithms is the qubit lattice algorithm (QLA) [1-21].
After identifying an appropriate set of qubits, QLA proceeds to define a unitary set of interleaved non-commuting collision-streaming operators which acts on this basis set of qubits so as to perturbatively recover the classical physics of interest.

The entanglement of qubits is at the essence of an efficient quantum algorithm.  A maximally entangled 2-qubit state is known as a Bell state [22].  Now the Hilbert space of a 2-qubit basis consists of the states $\{|00\rangle,|01\rangle,|10\rangle,|11\rangle\}$.  Consider the collision operator

\begin{equation}
C=\left[\begin{array}{cc}
\cos \theta & \sin \theta \\
-\sin \theta & \cos \theta
\end{array}\right]
\end{equation}
acting on the subspace $\{|00\rangle,|11\rangle\}$.  The most general tensor product state that can be generated from the qubit states 
$\{a_{0} |0\rangle + a_{1}  |1\rangle\}$ , and $\{b_{0} |0\rangle + b_{1}  |1\rangle\}$  is 

\begin{equation}
a_{0} b_{0}  |00\rangle + a_{0} b_{1}  |01\rangle + a_{1} b_{0} |10\rangle+a_{1} b_{1} |11\rangle
\end{equation}

Now consider the so-called Bell state
\begin{equation}
B_{+}=\frac{|00\rangle+|11\rangle}{\sqrt{2}}  .
\end{equation}
This state cannot be recovered from the tensor-product state of the 2 qubits, Eq. (2).  Indeed, to eliminate the $ |01\rangle$ state from Eq. (2) one requires either $a_0 = 0$ or
$b_1 = 0$ - and this would eliminate either the state $|00\rangle$ or the state $|11\rangle$.  States that can not be recovered from tensor product states are called
entangled states.  The entangled Bell state Eq, (3) is obtained usingthe collision operator $C$, Eq. (1), with angle $\theta = \pi/4$.

It is simplest to develop a QLA for the two curl-Maxwell equations, treating the divergence equations as initial constraints on the electromagnetic fields $\mathbf{E, H}$.  We shall do this in a Hermitian tensor dielectric medium, and comment on the discreteness effects on the time evolution of $\nabla \cdot \mathbf{B}$. 
In Sec. 2 we shall see that in an inhomogeneous medium, the electromagnetic basis set $(\mathbf{E, B})$ cannot lead to a unitary evolution of the two curl Maxwell equations.  However, a Dyson map is introduced that will map the basis $(\mathbf{E, B})$  into the basis $(n_x E_x, n_y E_y. n_z E_z, \mathbf{B})$ resulting in a fully unitary evolution for this basis set [23].   Here we have transformed to principal axes making the dielectric tensor diagonal with $\epsilon_i = n^2_i , i = x, y, z$.  
The more familiar complex Riemann-Silberstein-Weber basis $F_i^{\pm} = (n_i E_i \pm i B_i)$ is immediately generated from the real basis $(n_i E_i, B_i)$ by a unitary transformation so that this will also lead to a unitary time evolution representation.

In Sec. 2 we will develop a QLA for the solution of 2D Maxwell equations in a tensor Hermitian dielectric medium.  All our previous Maxwell QLA [16-18, 21] were restricted to scalar dielectrics.  We will present a simplified discussion of the Dyson map [23] that will permit us to transform from a non-unitary to unitary basis for the representation of 
the two curl equations of Maxwell.  
For  these continuum qubit partial differential equations we will generate in Sec. 3  a discrete QLA for tensor dielectric media that recovers the desired equations to second order perturbation.  While the collide-stream operator sequence of QLA  is fully unitary, the external potential operators required to recover the derivatives of the refractive indices in Maxwell equations are not.  However these non-unitary matrices are very sparse and should be amenable to some unitary approximate representation.   
The role of the perturbation parameter $\delta$ introduced in the QLA for Maxwell equations is quite subtle.  One important test of the QLA is
the conservation of electromagnetic energy density.  This will be seen to be very well satisfied, as $\delta \rightarrow 0$.  
In Sec. 4 we present some 2D QLA simulations for a 1D Gaussian electromagnetic pulse scattering from an anisotropic dielectric localized object - showing results for both polarizations.  Finally, in Sec. 5 we summaries the results of this paper.

\section{A Unitary Representation of the two curl Maxwell Equations}
\subsection{Scalar dielectric medium}
First, consider a simple dielectric non-magnetic medium with the constitutive equations

\begin{equation}
\mathbf{D}=\epsilon \mathbf{E}, \quad \mathbf{B}=\mu_{0} \mathbf{H} .
\end{equation}

It is convenient to define $\mathbf{u}=(\mathbf{E}, \mathbf{H})^{\mathbf{T}}$ as the fundamental fields, and $\mathbf{d}=(\mathbf{D}, \mathbf{B})^{\mathbf{T}}$ the derived fields.  Eq. (4), in matrix form, is

\begin{equation}
\mathbf{d}=\mathbf{W u} .
\end{equation}

\noindent  $\mathbf{W}$ is a Hermitian $6 \times 6$ matrix

\begin{equation}
\mathbf{W}=\left[\begin{array}{cc}
\epsilon \mathbf{I}_{3 \times 3} & 0_{3 \times 3} \\
0_{3 \times 3}& \mu_{0} \mathbf{I}_{3 \times 3}
\end{array}\right]  .
\end{equation}

\noindent $\mathbf{I}_{3 \times 3}$ is the $3 \times 3$ identity matrix. and the superscript $\mathbf{T}$ is the transpose operator. The curl-curl  Maxwell equations $\nabla \times \mathbf{E}=-\partial \mathbf{B} / \partial t$, and $\nabla \times \mathbf{H}=\partial \mathbf{D} / \partial t$ can then be written

\begin{equation}
i \frac{\partial \mathbf{d}}{\partial t}=\mathbf{M u}
\end{equation}

\noindent where, under standard boundary conditions, the curl-matrix operator $\mathbf{M}$ is Hermitian :

\begin{equation}
\mathbf{M}=\left[\begin{array}{cc}
0_{3 \times 3} & i \nabla \times \\
-i \nabla \times & 0_{3 \times 3}
\end{array}\right]  .
\end{equation}

\noindent Now $\mathbf{W}$ is invertible, so that Eq. (7) can finally be written in terms of the basic electromagnetic fields $\mathbf{u}=(\mathbf{E}, \mathbf{H})$

\begin{equation}
i \frac{\partial \mathbf{u}}{\partial t}=\mathbf{W}^{-\mathbf{1}} \mathbf{M} \mathbf{u}
\end{equation}

\subsubsection{inhomogeneous scalar dielectric media}
We immediately note that for inhomogeneous dielectric media, $\mathbf{W^{-1}}$ will not commute with $\mathbf{M}$.
Thus Eq. (9)  will not yield unitary evolution for the fields $\mathbf{u}=(\mathbf{E}, \mathbf{H})^{\mathbf{T}}$ .
However Koukoutsis et. al. [23] have shown how to determine a Dyson map from the fields $\mathbf{u}$ to a new field representation $\mathbf{U}$ such that the resultant representation in terms of the new field $\mathbf{U}$ will result in a unitary evolution. 
In particular,  the Dyson map [23]

\begin{equation}
\mathrm{U}=\mathrm{W}^{1 / 2} \mathrm{u}
\end{equation}

\noindent yields a unitary evolution equation for $\mathbf{U}$ :

\begin{equation}
i \frac{\partial \mathbf{U}}{\partial t}=\mathbf{W}^{-1 / 2} \mathbf{M} \mathbf{W}^{-1 / 2} \mathbf{U}
\end{equation}

\noindent since now the matrix operator $\mathbf{W}^{-\mathbf{1} / \mathbf{2}} \mathbf{M} \mathbf{W}^{-\mathbf{1} / \mathbf{2}}$ is indeed Hermitian.

Explicitly, the $\mathbf{U}$ vector for non-magnetic materials,is just

\begin{equation}
\mathbf{U}=\left(\epsilon^{1 / 2} \mathbf{E}, \mu_{0}^{1 / 2} \mathbf{H}\right)^{T}
\end{equation}

\noindent This can be rotated into the RWS unitary representation by the unitary matrix

\begin{equation}
\mathbf{L}=\frac{1}{\sqrt{2}}\left[\begin{array}{cc}
I_{3 \times 3} & i I_{3 \times 3} \\
I_{3 \times 3} & -i I_{3 \times 3}
\end{array}\right]
\end{equation}

\noindent yielding $\mathbf{U}_{\mathbf{R S W}}=\mathbf{L U}$ with

\begin{equation}
\mathbf{U}_{\mathbf{R S W}}=\frac{1}{\sqrt{2}}\left[\begin{array}{c}
\epsilon^{1 / 2} \mathbf{E}+i \mu_{0}^{1 / 2} \mathbf{H} \\
\epsilon^{1 / 2} \mathbf{E}-i \mu_{0}^{1 / 2} \mathbf{H}
\end{array}\right] .
\end{equation}

\subsection{Inhomogeneous tensor dielectric media}

The theory can be immediately extended to diagonal tensor dielectric media, with (assuming non-magnetic materials) the 6-qubit representation $\mathbf{Q}$ of the field
\begin{equation}
\mathbf{U}=\left(n_x E_x, n_y E_y, n_z E_z, \mu_{0}^{1 / 2} \mathbf{H}\right)^{T}  \equiv \mathbf{Q}   .
\end{equation}
\noindent $(n_x , n_y, n_z)$ is the vector (diagonal) refractive index, with $\epsilon_x = n_x^2$ ...  .

The explicit unitary representation of the Maxwell equations for 2D x-y spatially dependent fields written in terms of the 6-$\mathbf{Q}$ qubit components are
\begin{equation}
\begin{aligned}
\frac{\partial q_0}{\partial t} = \frac{1}{n_x} \frac{\partial q_5}{\partial y} , \qquad
\frac{\partial q_1}{\partial t} = - \frac{1}{n_y} \frac{\partial q_5}{\partial x} , \qquad
\frac{\partial q_2}{\partial t} =  \frac{1}{n_z} \left[ \frac{\partial q_4}{\partial x} -\frac{\partial q_3}{\partial y} \right] \\
\frac{\partial q_3}{\partial t} = - \frac{\partial (q_2/n_z)}{\partial y} , \qquad
\frac{\partial q_4}{\partial t} = \frac{\partial (q_2/n_z)}{\partial x} , \qquad
\frac{\partial q_5}{\partial t} = - \frac{\partial (q_1/n_y)}{\partial x}  + \frac{\partial (q_0/n_x)}{\partial y} 
\end{aligned}
\end{equation}

\section{A Qubit Lattice Representation for 2D Tensor Dielectric Media}

We develop a QLA for the unitary system Eq. (16) by determining unitary collision and streaming operators that recover the derivatives $\partial q_i/\partial t, \partial q_j/\partial x$ and $\partial q_j/\partial y$. ($i, j = 1..6$).  Our finite difference scheme is to recover Eq. (16) to second order in a perturbation parameter $\delta$, where the
spatial lattice spacing is defined to be $O(\delta)$.  To recover the partial derivatives on the 6-qubit $\mathbf{Q}$ in the $x-$direction,  we consider the unitary collision entangling operator

\begin{equation}
C_X=\left[\begin{array}{cccccc}
1 & 0 & 0 & 0& 0& 0 \\
0 & cos \,\theta_1 & 0 & 0 & 0 & - sin\,\theta_1 \\
0 & 0 & cos\,  \theta_2 & 0 & - sin \,\theta_2 & 0 \\
0 & 0 & 0 & 1 &  0 & 0  \\
0 & 0 & sin\,\theta_2 & 0 & cos\, \theta_2 & 0 \\
0 & sin\, \theta_1 & 0 & 0 & 0 & cos \,\theta_1
\end{array}\right]
\end{equation}
where we shall need two collision angles $\theta_1$ and $\theta_2$.  The unitary streaming operators will be of the form $S^{+x}_{14}$  which shifts qubits $q_1$ and $q_4$ one lattice unit $\delta$ in the $+x-$direction, while leaving the other 4 qubit components invariant.  The final unitary collide-stream sequence in the x-direction is
\begin{equation}
\mathbf{U_X} = S^{+x}_{25}.C_X^\dag . S^{-x}_{25}.C_X. S^{-x}_{14}.C_X^\dag . S^{+x}_{14}.C_X .S^{-x}_{25}.C_X . S^{+x}_{25}.C_X^\dag. S^{+x}_{14}.C_X . S^{-x}_{14}.C_X^\dag 
\end{equation}.

Similarly for the y-direction, the corresponding unitary collision entangling operator is
\begin{equation}
C_Y=\left[\begin{array}{cccccc}
cos\, \theta_0 & 0 & 0 & 0& 0& sin\, \theta_0 \\
0 & 1 & 0 & 0 & 0 & 0 \\
0 & 0 & cos\,  \theta_2 & 0 &  sin \,\theta_2 & 0 \\
0 & 0 & -sin\,\theta_2 & cos \,\theta_2 &  0 & 0  \\
0 & 0 & 0& 0 & 1 & 0 \\
-sin \,\theta_0& 0 & 0 & 0 & 0 & cos \,\theta_0
\end{array}\right]   ,
\end{equation}
\noindent and the corresponding unitary collide-stream sequence in the y-direction
\begin{equation}
\mathbf{U_Y} = S^{+y}_{25}.C_Y^\dag . S^{-y}_{25}.C_Y. S^{-y}_{03}.C_Y^\dag . S^{+y}_{03}.C_Y .S^{-y}_{25}.C_Y . S^{+y}_{25}.C_Y^\dag. S^{+y}_{03}.C_Y . S^{-y}_{03}.C_Y^\dag 
\end{equation}
We will discuss the specific collision angles $\theta_0, \theta_1$ and $\theta_2$ after introducing the external potential operators.

The terms that remain to be recovered by the QLA are the spatial derivatives on the refractive index components $\partial n_i/\partial x$ and $\partial n_i/\partial y$.  These terms will be recovered by the following (non-unitary) sparse external potential operators: 
\begin{equation}
V_X=\left[\begin{array}{cccccc}
1 & 0 & 0 & 0& 0& 0 \\
0 & 1 & 0 & 0 & 0 & 0\\
0 & 0 & 1 & 0 &0& 0 \\
0 & 0 & 0 & 1 &  0 & 0  \\
0 & 0 &- sin\,\beta_2 & 0 & cos\, \beta_2 & 0 \\
0 & sin\, \beta_0 & 0 & 0 & 0 & cos \,\beta_0
\end{array}\right]
\end{equation}
\noindent and
\begin{equation}
V_Y=\left[\begin{array}{cccccc}
1 & 0 & 0 & 0& 0& o \\
0 & 1 & 0 & 0 & 0 & 0\\
0 & 0 & 1 & 0 &0& 0 \\
0 & 0 & \cos\, \beta_3& \sin \, \beta_3 &  0 & 0  \\
0 & 0 & 0 & 0 & 1 & 0 \\
-sin\, \beta_1 & 0 & 0 & 0 & 0 & cos \,\beta_1
\end{array}\right]
\end{equation}
\noindent  for particular angles $\beta_0 \, .. \, \beta_3$.  

Thus one possible QLA algorithm that advances the 6-qubit $\mathbf{Q}$ from time $t$ to time $t+\Delta t$ is
\begin{equation}
\mathbf{Q}(t+\Delta t)  = V_Y . V_X. \mathbf{U_Y} . \mathbf{U_X} . \mathbf{Q}(t)
\end{equation}
Indeed, using Mathematica, one can show that with the collision angles
\begin{equation}
\theta_0 = \frac{\delta}{4 n_x} \quad , \qquad \theta_1 = \frac{\delta}{4 n_y} \quad , \qquad  \theta_2 = \frac{\delta}{4 n_z} ,
\end{equation}
\noindent and
\begin{equation}
\beta_0 = \delta^2 \frac{\partial n_y/\partial x}{n^2_y} \quad , \quad \beta_1 = \delta^2 \frac{\partial n_x/\partial y}{n^2_x}  \quad , \quad \beta_2 = \delta^2 \frac{\partial n_z/\partial x}{n^2_z} \quad , \quad \beta_3 = \delta^2 \frac{\partial n_z/\partial y}{n^2_z}
\end{equation}
\noindent we will have a second order QLA representation of the 2D Maxwell continuum equations
\begin{equation}
\begin{aligned}
\frac{\partial q_0}{\partial t} =\frac{ \delta^2}{ \Delta t} \frac{1}{n_x} \frac{\partial q_5}{\partial y} + O(\frac{\delta^4}{\Delta t})\\
\frac{\partial q_1}{\partial t} = -\frac{ \delta^2}{  \Delta t}  \frac{1}{n_y} \frac{\partial q_5}{\partial x} + O(\frac{\delta^4}{\Delta t})\\
\frac{\partial q_2}{\partial t} =\frac{  \delta^2}{ \Delta t}   \frac{1}{n_z} \left[ \frac{\partial q_4}{\partial x} -\frac{\partial q_3}{\partial y} \right] + O(\frac{\delta^4}{\Delta t})\\
\frac{\partial q_3}{\partial t} = -\frac{ \delta^2}{ \Delta t}  \left[ \frac{1}{n_z} \frac{\partial q_2}{\partial y}  - \frac{\partial n_z/\partial y}{n_z^2} q_2 \right]+ O(\frac{\delta^4}{\Delta t})\\
\frac{\partial q_4}{\partial t} = \frac{ \delta^2}{ \Delta t}  \left[ \frac{1}{n_z} \frac{\partial q_2}{\partial x}  - \frac{\partial n_z/\partial x}{n_z^2} q_2 \right]+ O(\frac{\delta^4}{\Delta t})\\
\frac{\partial q_5}{\partial t} = \frac{\delta^2}{ \Delta t } \left[ -\frac{1}{n_y} \frac{\partial q_1}{\partial x}  +\frac{\partial n_y/\partial x}{n_y^2} q_1 + \frac{1}{n_x} \frac{\partial q_0}{\partial y}  -\frac{\partial n_x/\partial y}{n_x^2} q_0  \right] + O(\frac{\delta^4}{\Delta t})\
\end{aligned}
\end{equation}
\noindent under diffusion ordering, $\Delta t \approx \delta^2$.

\subsection{Conservation of Instantaneous Total Electromagnetic Energy in QLA Simulations}
It is important to monitor the conservation of energy in the QLA, particularly since our current QLA is not fully unitary.
The normalized total electromagnetic energy for a square lattice domain of length $L$ is $\mathcal{E}(t)$
\begin{equation}
\mathcal{E}(t) = \frac{1}{L^2} \int_0^L \int_0^L dx dy \left[ n_x^2 E_x^2 + n_y^2 E_y^2 + n_z^2 E_z^2 + \mathbf{B}^2 \right] =  \frac{1}{L^2}  \int_0^L \int_0^L dx dy \mathbf{Q} \cdot \mathbf{Q} \quad ,
\end{equation}  
\noindent In our QLA simulations, we will consider the scattering of a 1D Gaussian pulse propagating in the $x-$direction,
and scattering from a localized tensor 2D dielectric object in the $x-y$ plane.  We choose $L$ to be significantly greater than 
the dielectric object so that for $y \approx 0$, and for $y \approx L$ the electromagnetic fields there will be that of the 1D Gaussian
pulse yielding a Poynting vector $\mathbf{E} \cross \mathbf{B}$ in the $\hat{\mathbf{x}}$.
Thus the contribution to the Poynting flux $\oint_C \, \mathbf{E} \cross \mathbf{B} \, \cdot d \mathbf{\ell}$  
on $y=0$ and on $y=L$ is zero.
In our time evolution QLA simulations, we integrate only to $t < t_{max}$ so that there are no fields generated on the sides
$x=0$ and $x=L$.  Thus, in our QLA simulations we have set up parameters such that the total electromagnetic energy 
$\mathcal{E}(t) = const.$, Eq. (27), for $t < t_{max}$.

$\mathcal{E}(t)$ is nothing but the norm of $\mathbf{Q}-$qubits , and will be exactly conserved in a fully unitary QLA.
One must also be careful  in the ordering of the external potential angles, 
Eq. (25) :  they must be $O(\delta^2)$ in order to recover Maxwell equations.

While we will discuss in detail in Sect. 4 our numerical QLA simulation of a 1D electromagnetic pulse scattering from a localized dielectric object  it is appropriate to discuss here some QLA simulation results for the total energy.  Since QLA, Eq. (23) is a perturbation theory, it will recover the 2D Maxwell equations as $\delta \rightarrow 0$.  For $\delta = 0.3$, we find the following time variation in the total energy $\mathcal{E}(t)$ in Fig. 1a. $t_{max} = 20, 000$ lattice time steps.   
 \begin{figure}[!h!p!b!t] \ 
\begin{center}
\includegraphics[width=3.2in]{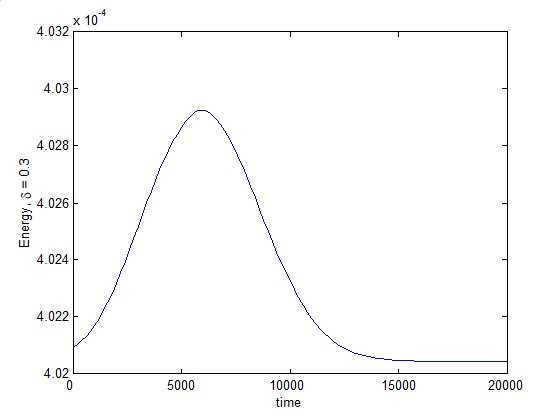}
\includegraphics[width=3.2in]{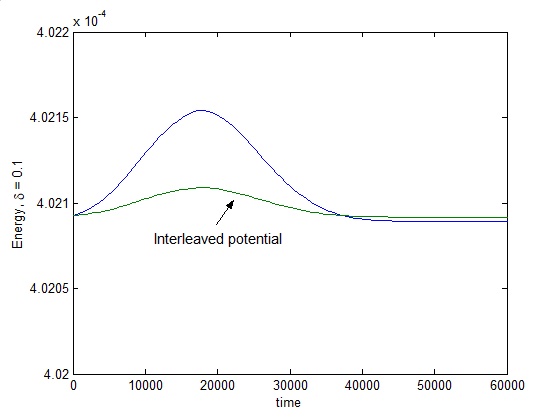}
(a)  $\mathcal{E}$(t) , $\delta=0.3$  , \qquad \qquad  \qquad  (b)  $\mathcal{E}$(t) , $\delta=0.1$ 
\caption{ The instantaneous total electromagnetic energy $\mathcal{E}(t)$, Eq. (27), for various values of the perturbation parameter $\delta$ :  (a)  $\delta = 0.3$, (b)  $\delta = 0.1$.  A
more accurate QLA results from interleaving the external potentials with the unitary collide-stream operators. For $\delta = 0.01$, $\mathcal{E}(t)$ shows no variation on this scale, with 
variations in the 9th significant figure.
Lattice grid L = 8192.
}
\end{center}
\end{figure}

On lowering the perturbation parameter to $\delta = 0.1$ there is a nice reduction in the time variation of
$\mathcal{E}(t)$, Fig 1(b).  To reach the same physics $t_{max} = 60K$.  

However, if we interleave the external potential operators among the unitary collide-stream sequence (and similarly for the y-direction) in the form
\begin{equation}
\mathbf{V_X' U_X} = V_X' S^{+x}_{25}.C_X^\dag . S^{-x}_{25}.C_X. S^{-x}_{14}.C_X^\dag . S^{+x}_{14}.C_X .V_X'.S^{-x}_{25}.C_X . S^{+x}_{25}.C_X^\dag. S^{+x}_{14}.C_X . S^{-x}_{14}.C_X^\dag 
\end{equation}
\noindent (with the corresponding potential angle reduced by a factor of 2) we find $\mathcal{E}(t) \approx  const.$ for all times, see Fig. 1(b).  There is a further strong improvement in $\mathcal{E}(t) = const.$ for $\delta = 0.01$.  

\section{Scattering of a Polarized Pulse from an Anisotropic Dielectric Object}
We first consider a 1D Gaussian pulse propagating in a vacuum in the x-direction towards a localized anisotropic dielectric object, with diagonal tensor components which are conical in $n_z(x,y)$, and cylindrical in the $x$ and $y$ directions with $n_x(x,y) = n_y(x,y)$, Fig. 2
 \begin{figure}[!h!p!b!t] \ 
\begin{center}
\includegraphics[width=3.2in]{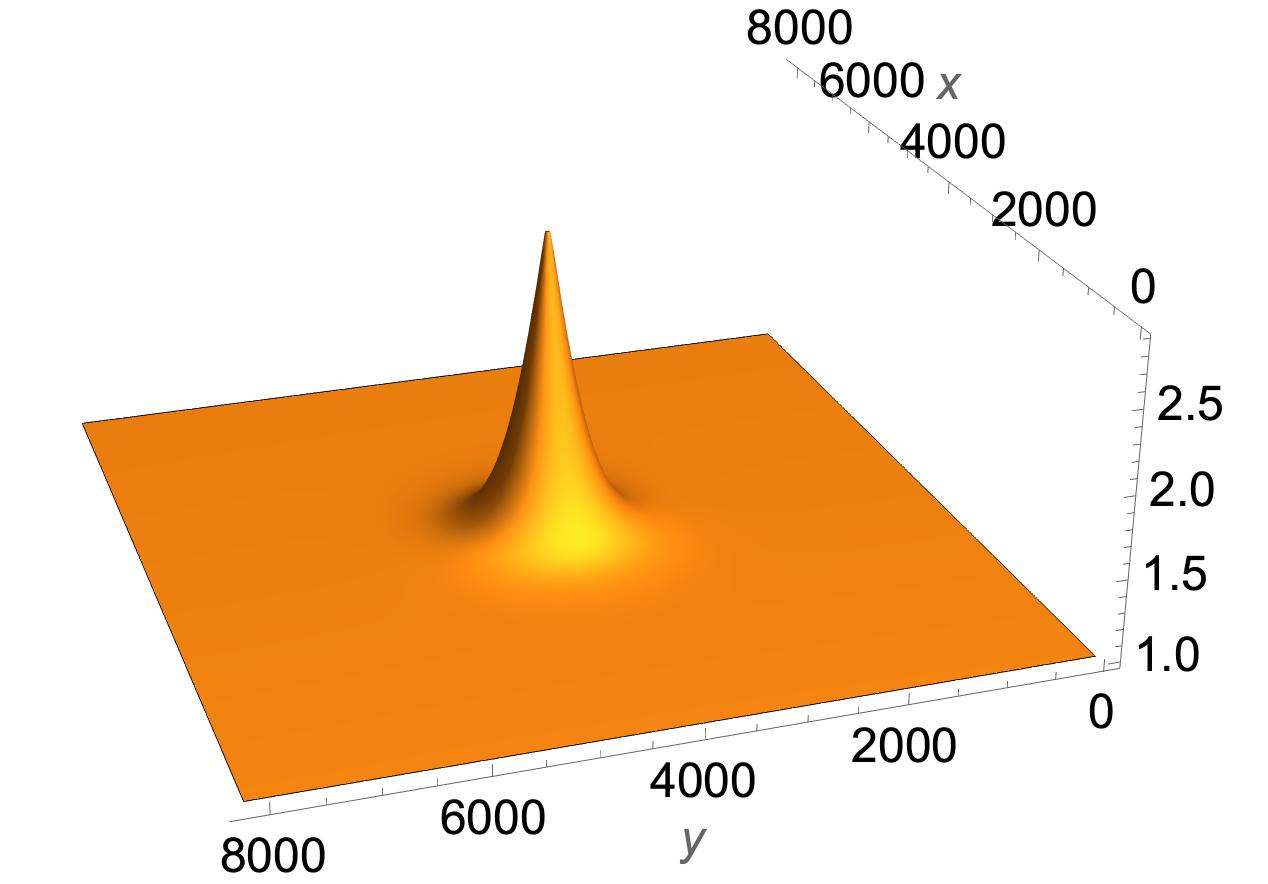}
\includegraphics[width=3.2in]{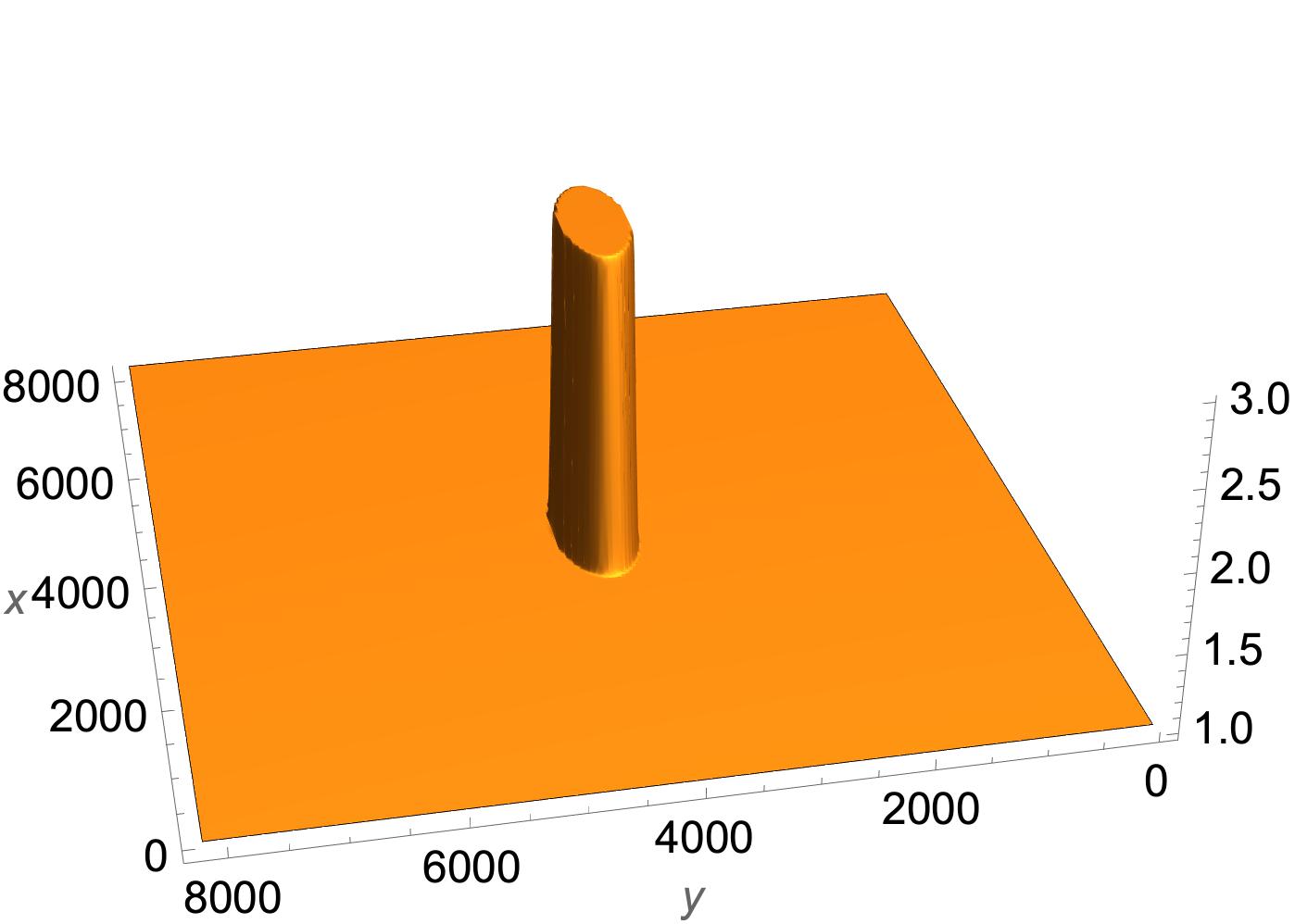}
(a) $n_z(x,y)$   \qquad ,  \qquad  \qquad  (b)  $n_x(x,y) = n_y(x,y)$ 
\caption{Anisotropic tensor dielectric :  (a)  conical in $n_z$, and (b)  cylindrical in $n_x = n_y$.  Initially, a 1D Gaussian pulse propagates in the $x$-direction, with either a polarization $E_z(x,t) < 0$ or a polarization $E_y(x,t)$ and scatters off this tensor dielectric object.  In the region away from the tensor dielectric object, we have a vacuum with $n_i = 1.0$.  In the dielectric, $n_{i,max} = 3.0$.  Lattice domain $8192^2$.
}
\end{center}
\end{figure}

\subsection{Scattering of 1D pulse with $E_z$ polarization}

When the 1D pulse with non-zero $E_z(x,t), B_y(x,t)$ fields starts to interact with the 2D tensor dielectric $\mathbf{n}(x,y)$, the scattered fields become 2D (see Fig. 3), with $B_y(x,y,t)$ dependence.  The QLA will then spontaneously generate a $B_x(x,y,t)$ field 
so that $\partial B_x/ \partial x  + \partial B_y/ \partial y \approx  0$.
 \begin{figure}[!h!p!b!t] \ 
\begin{center}
\includegraphics[width=3.2in]{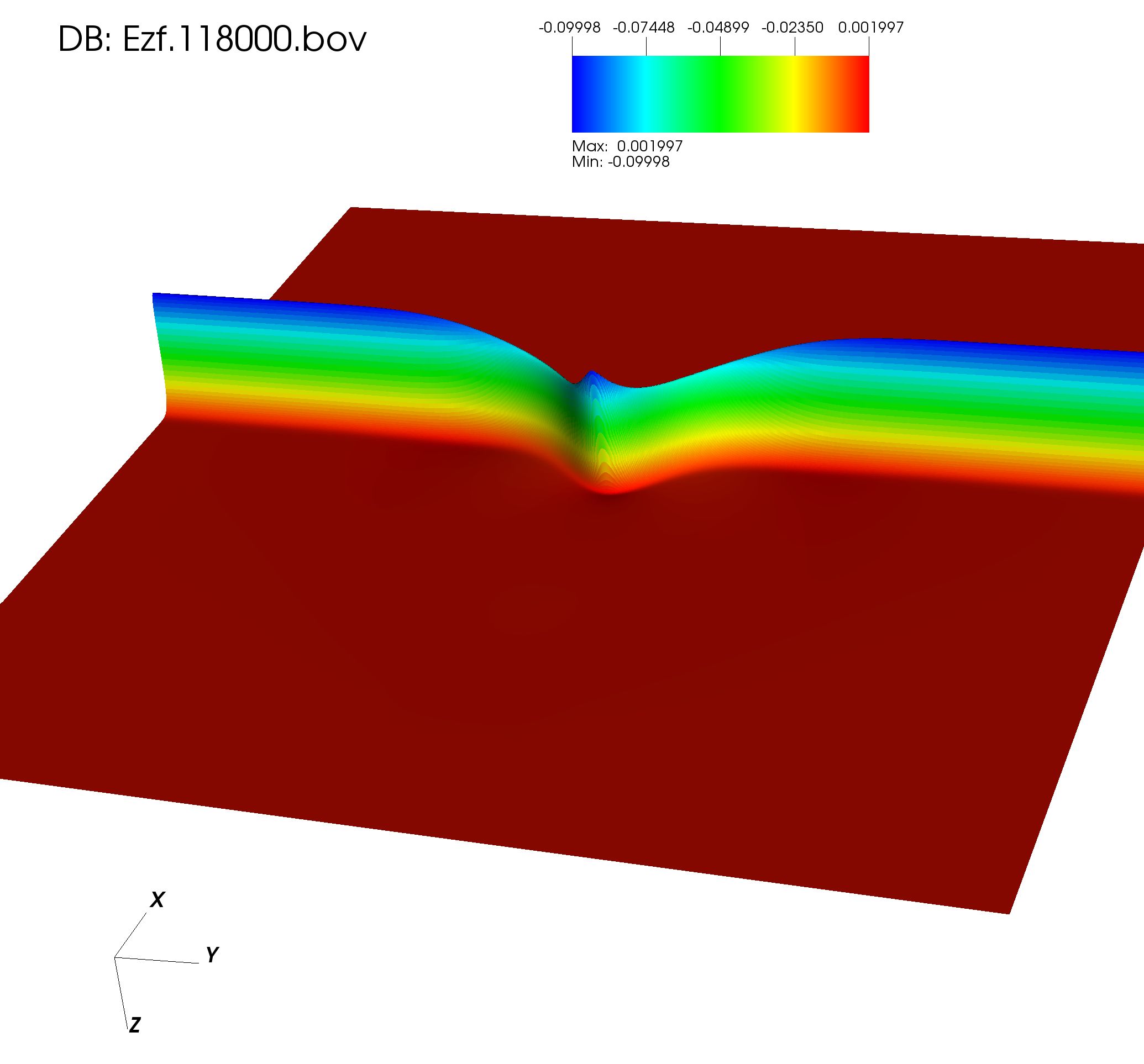}
\includegraphics[width=3.2in]{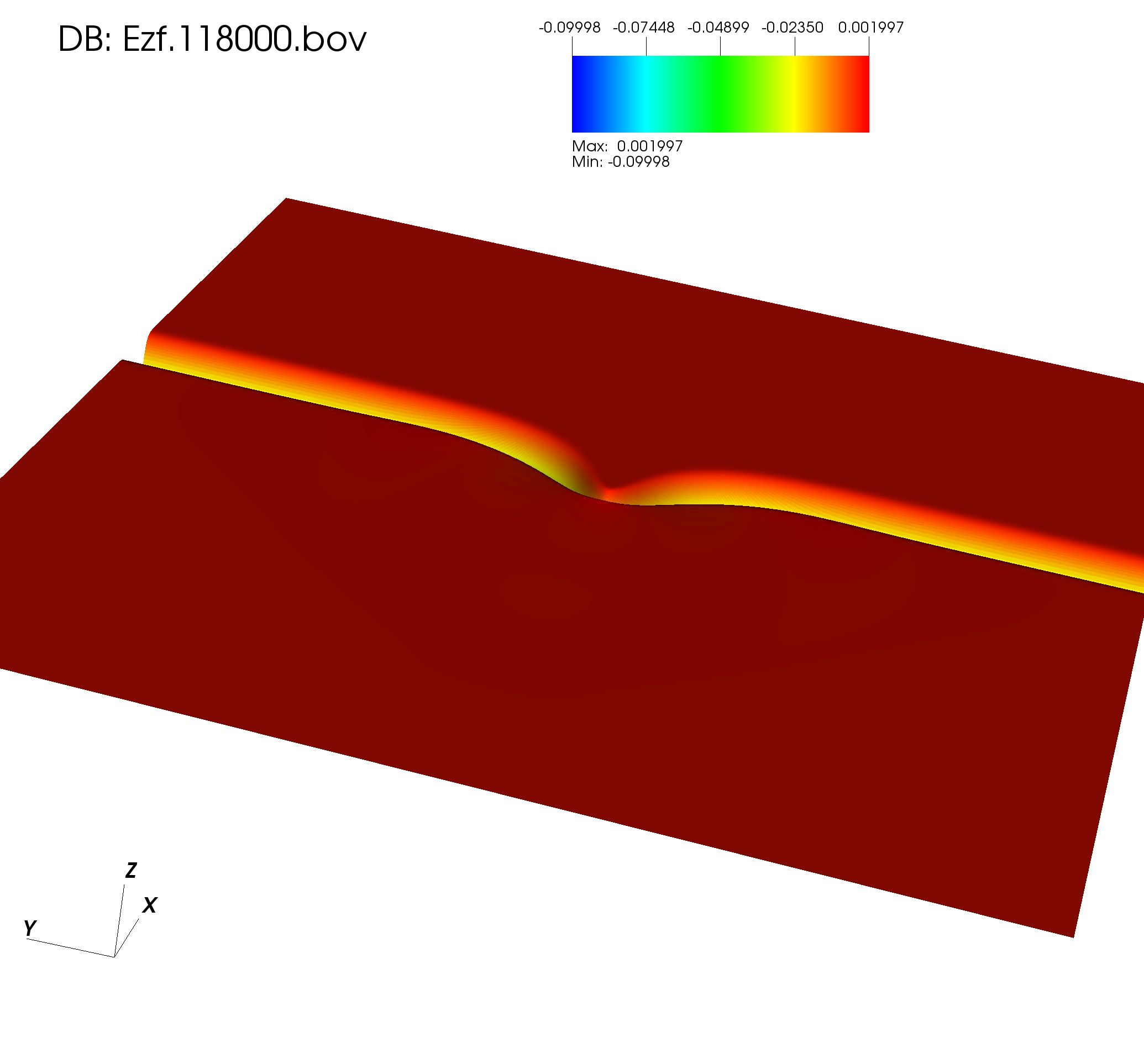}
(a) $E_z(x,y,t_0) < 0 $ at $t_0 = 18k$   \qquad ,  \qquad  \qquad  (b)  $E_z(x,y,t_0) > 0$ at $t_0 = 18k$
\caption{$E_z$ after interacting with the localized tensor dielectric.  Since the phase speed in the tensor dielectric is less than in the vacuum, the 2D structure in $E_z$ lags the rest of the 1D pulse that has not interacted with the localized dielectric object (Fig. 1).  The perspective
(b) is obtained from (a) by rotating by $\pi$ about the line $y=L/2$.
}
\end{center}
\end{figure}

Because of the relatively weak dielectric tensor gradients for a cone, there is very little reflection back into the vacuum of the incident $E_z$ field (Fig. 4).  There is a localized
transmitted $E_z$ within the dielectric.

 \begin{figure}[!h!p!b!t] \ 
\begin{center}
\includegraphics[width=3.2in]{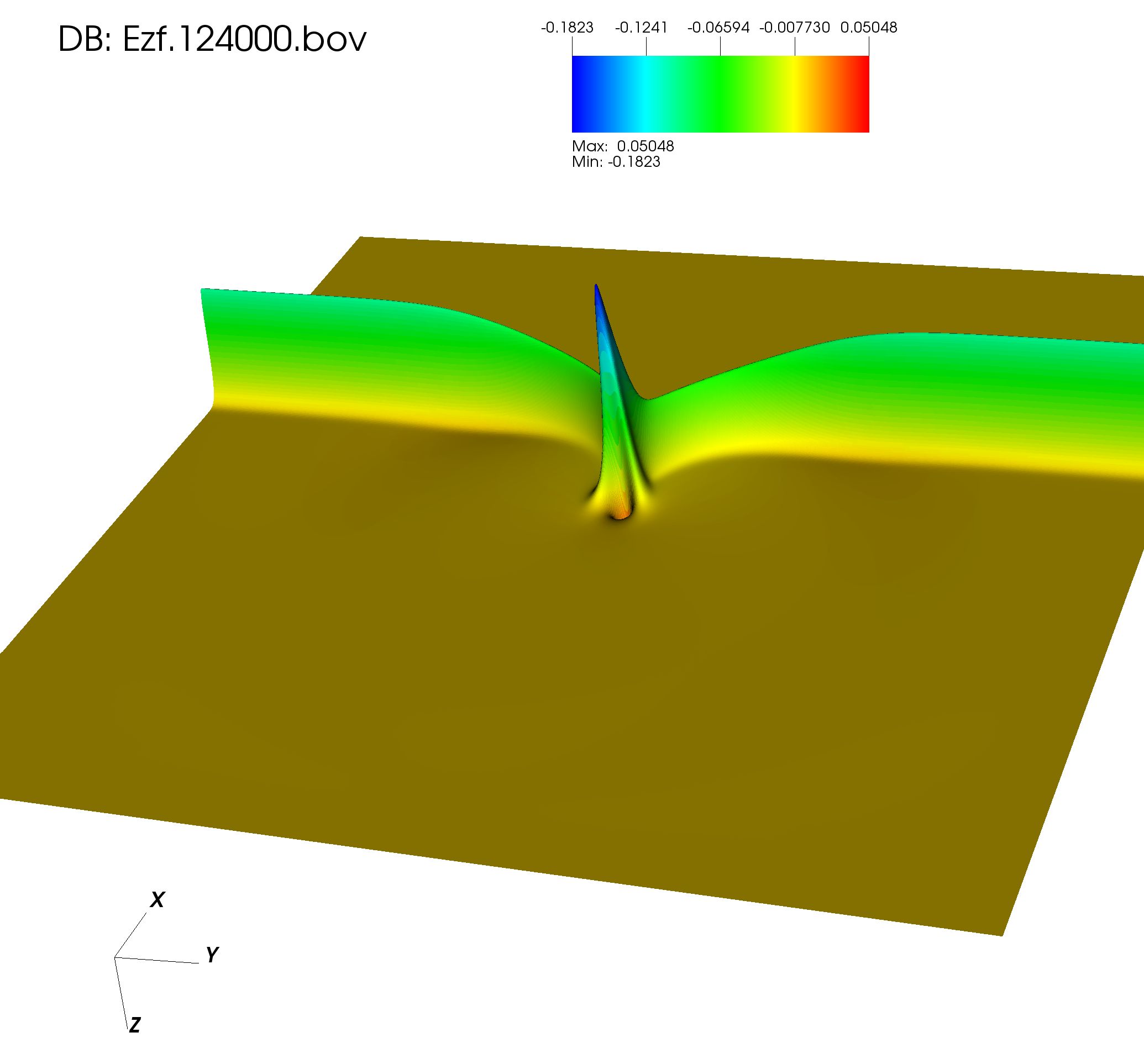}
\includegraphics[width=3.2in]{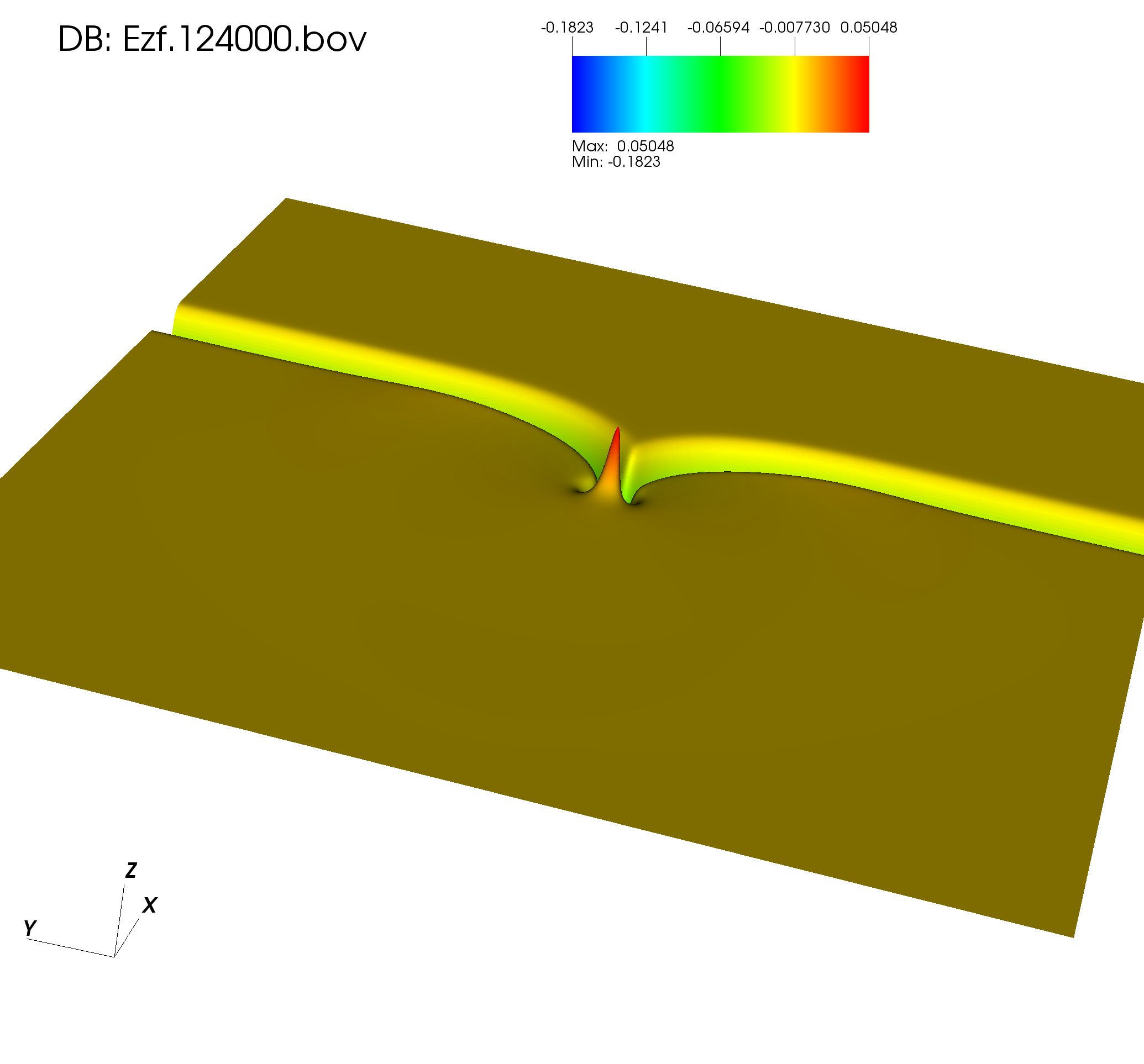}
(a) $E_z(x,y,t_1) < 0 $ at $t_1 = 24k$   \qquad ,  \qquad  \qquad  (b)  $E_z(x,y,t_1) > 0$ at $t_1 = 24k$
\caption{For early times, from the perspective of the tensor dielectric object the electromagnetic pulse within the dielectric is the $transmitted$ field and has a localized $E_z$ which becomes greater than the original $E_z$ in the vacuum region.  There is little $reflected$ field since $E_z$ will be predominantly interacting with the $n_z$ component of the tensor dielectric.  The perspective
(b) is obtained from (a) by rotating by $\pi$ about the line $y=L/2$.
}
\end{center}
\end{figure}

At $t = 36k$ we plot both the $E_z$ and the $B_y$ , Fig. 5-6.  Of considerable interest is the spontaneously generated $B_x(x,y,t)$ field so that $\nabla \cdot \mathbf {B} = 0$.
From Fig. 7 we see that $B_x$ has dipole structure, since the plane of the plot Fig. 7(b) for the field $B_x < 0$ is generated by rotating the plane through $\pi$ about the axis $x = L/2$

 \begin{figure}[!h!p!b!t] \ 
\begin{center}
\includegraphics[width=3.2in]{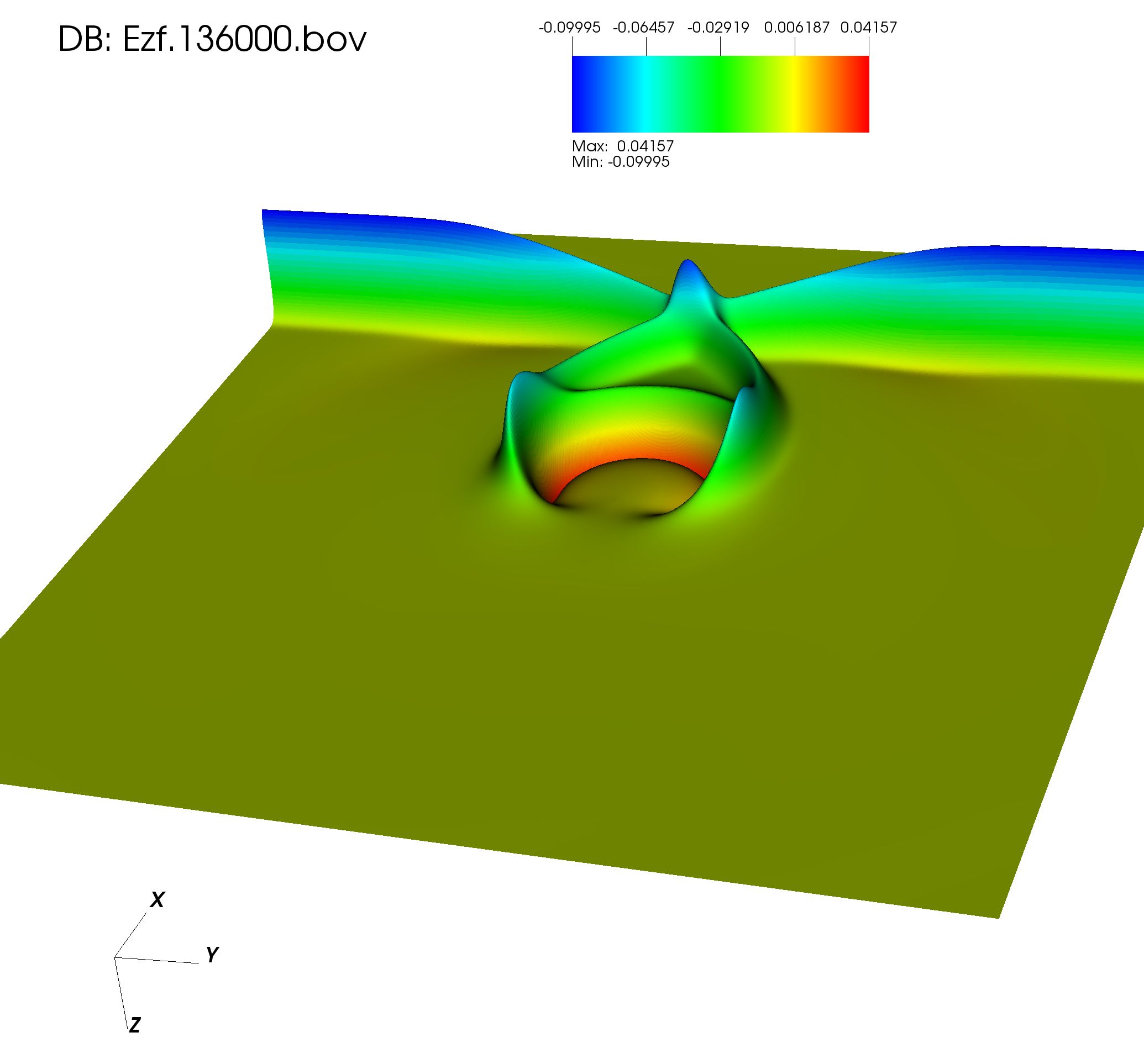}
\includegraphics[width=3.2in]{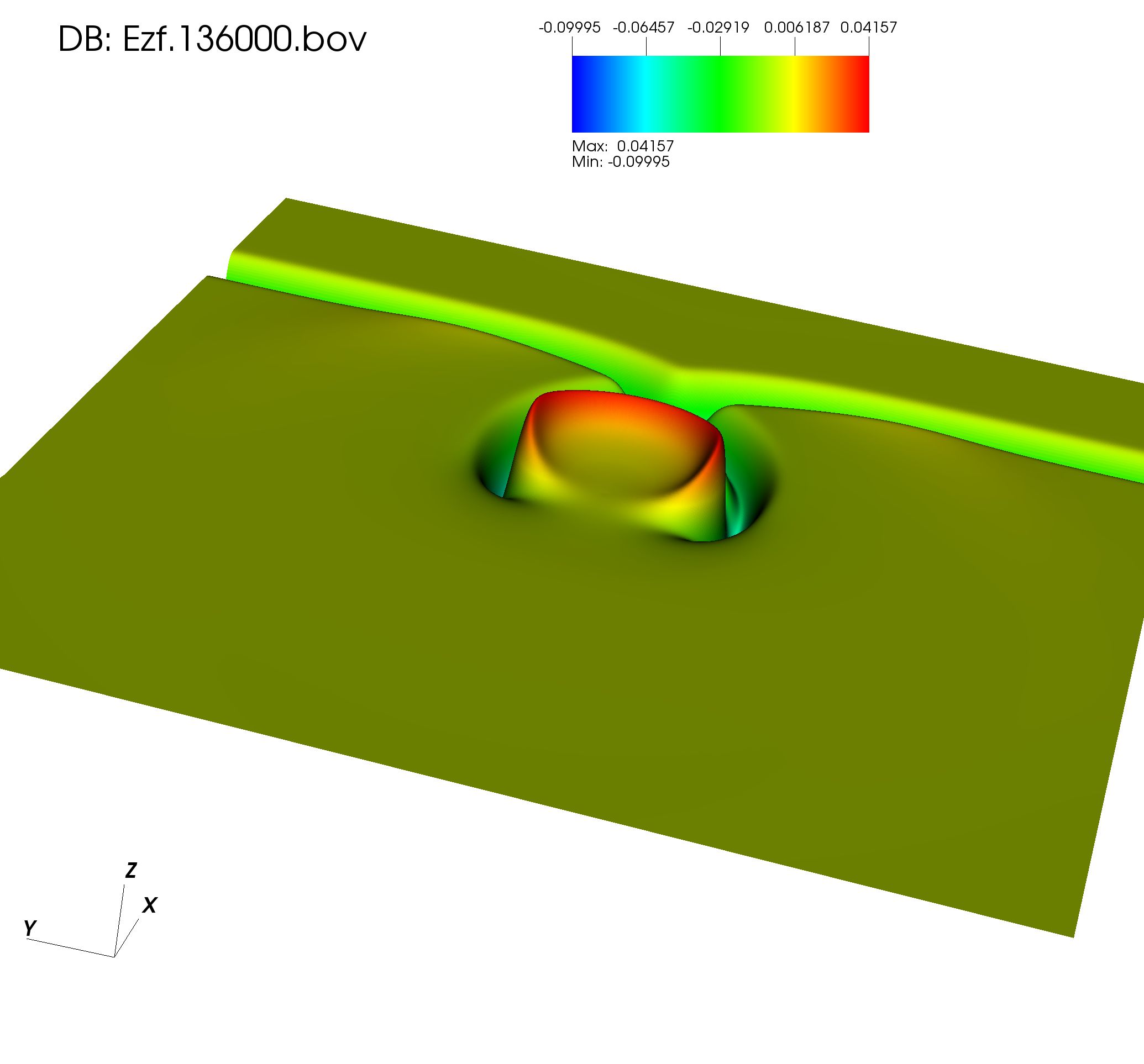}
(a) $E_z(x,y,t_2) < 0 $ at $t_2 = 36k$   \qquad ,  \qquad  \qquad  (b)  $E_z(x,y,t_2) > 0$ at $t_2 = 36k$
\caption{The $E_z$ field at a late stage of development.  The perspective
(b) is obtained from (a) by rotating by $\pi$ about the line $y=L/2$.
}
\end{center}
\end{figure}

 \begin{figure}[!h!p!b!t] \ 
\begin{center}
\includegraphics[width=3.2in]{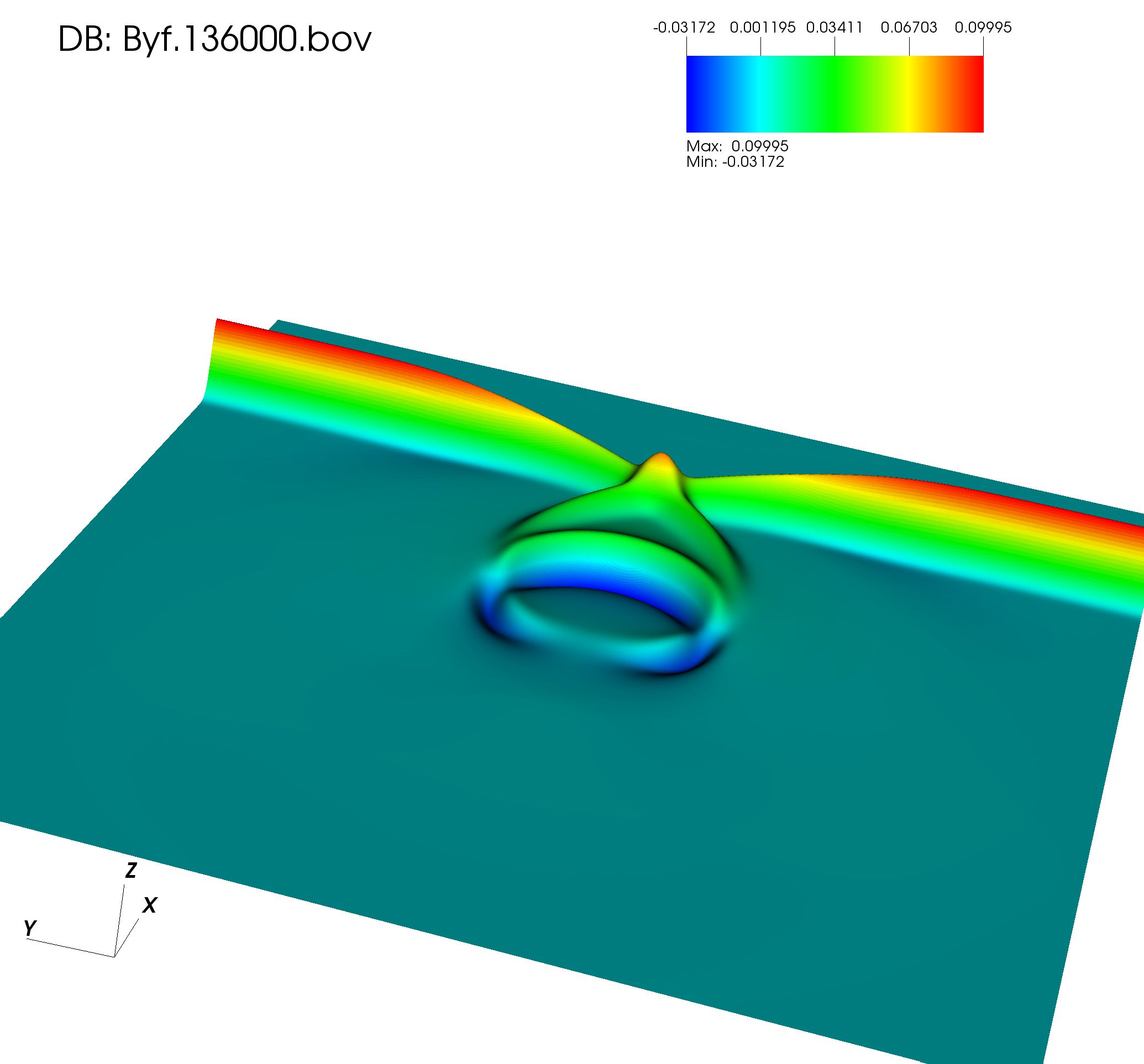}
\includegraphics[width=3.2in]{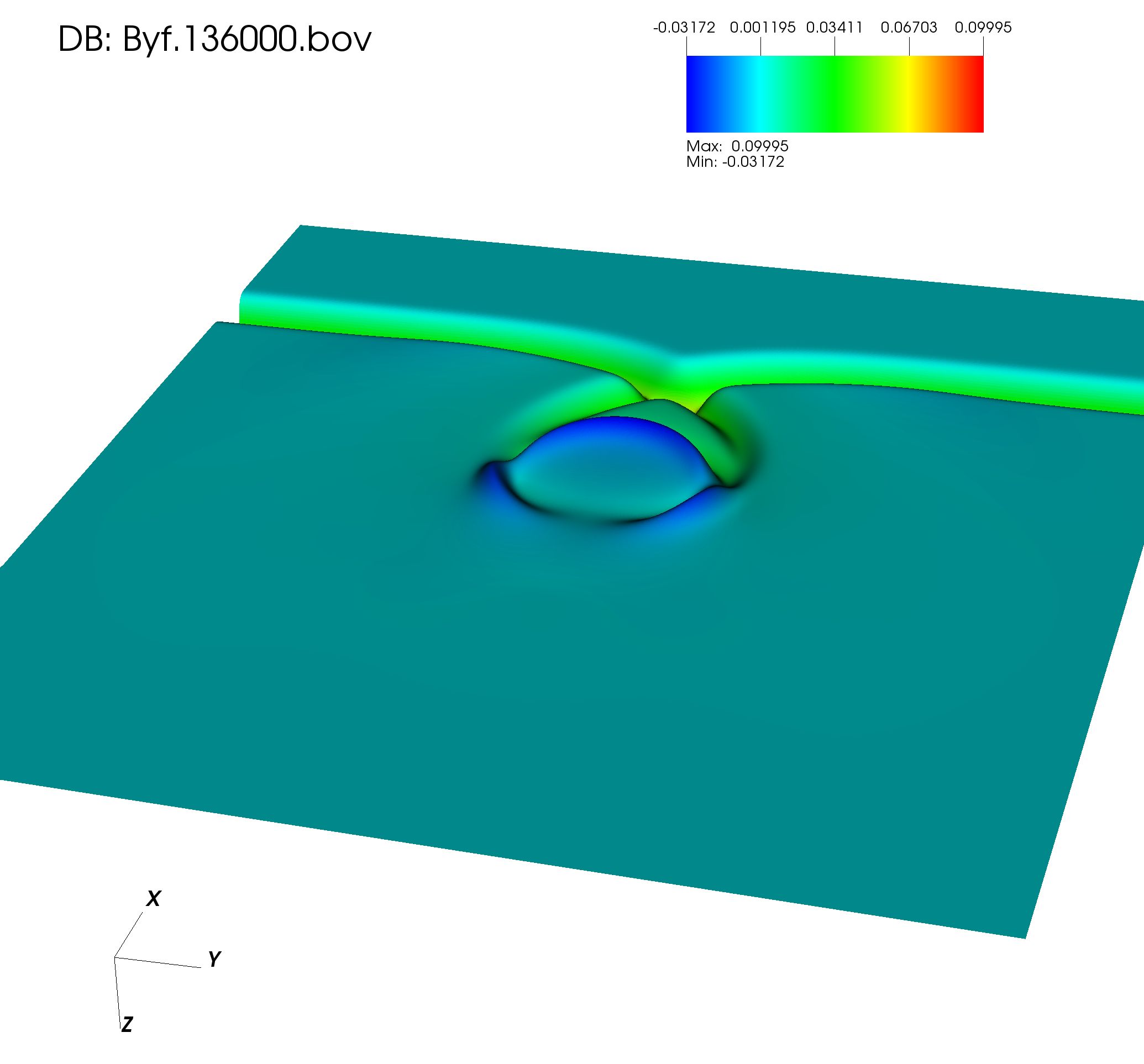}
(a) $B_y(x,y,t_0) > 0 $ at $t_0 = 36k$   \qquad ,  \qquad  \qquad  (b)  $B_y(x,y,t_0) < 0$ at $t_0 = 36k$
\caption{The corresponding $B_y$ field at time $t = 36k$ to the $E_z$ field in Fig. 5.  The perspective
(b) is obtained from (a) by rotating by $\pi$ about the line $y=L/2$.
}
\end{center}
\end{figure}

 \begin{figure}[!h!p!b!t] \ 
\begin{center}
\includegraphics[width=3.2in]{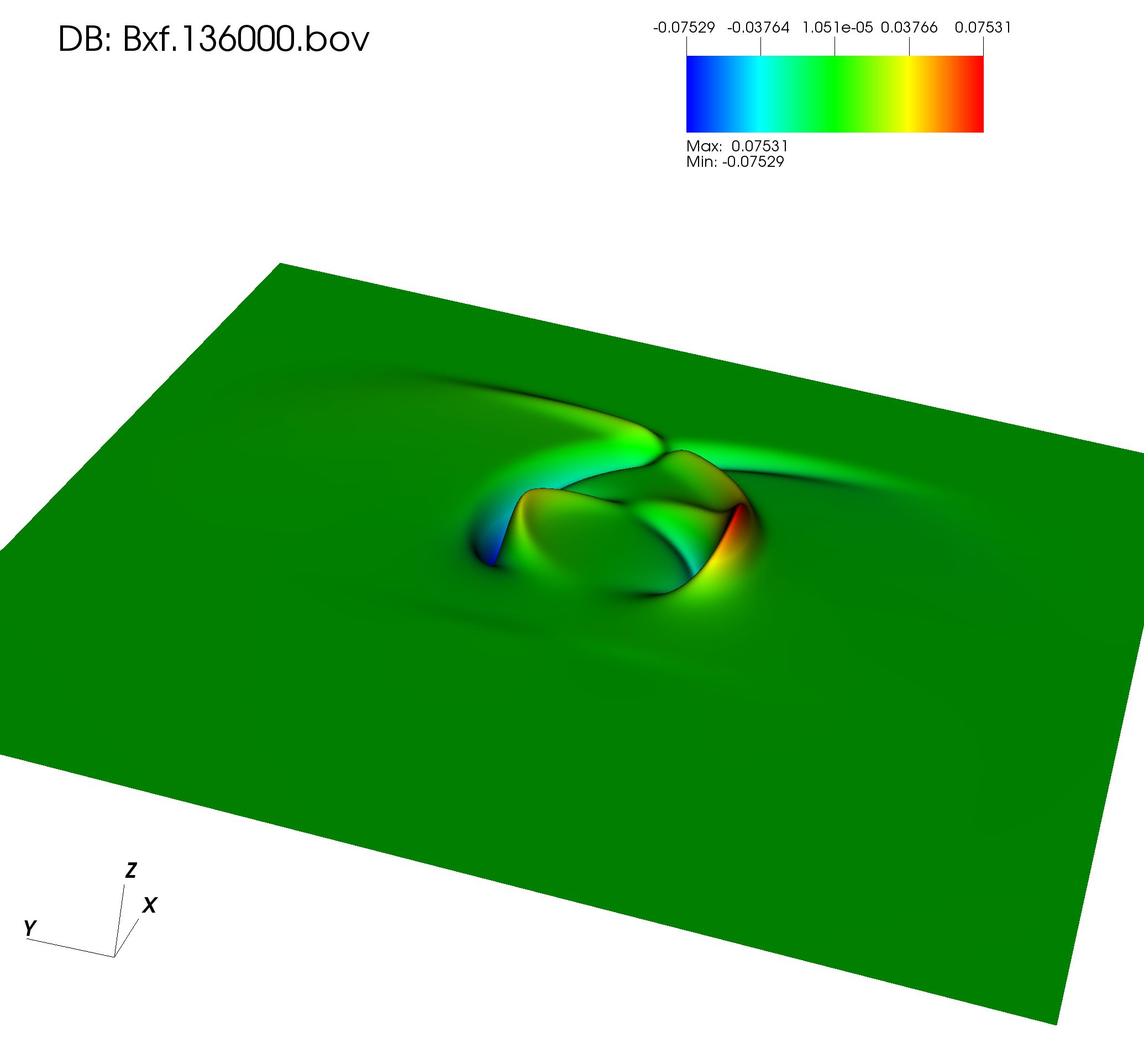}
\includegraphics[width=3.2in]{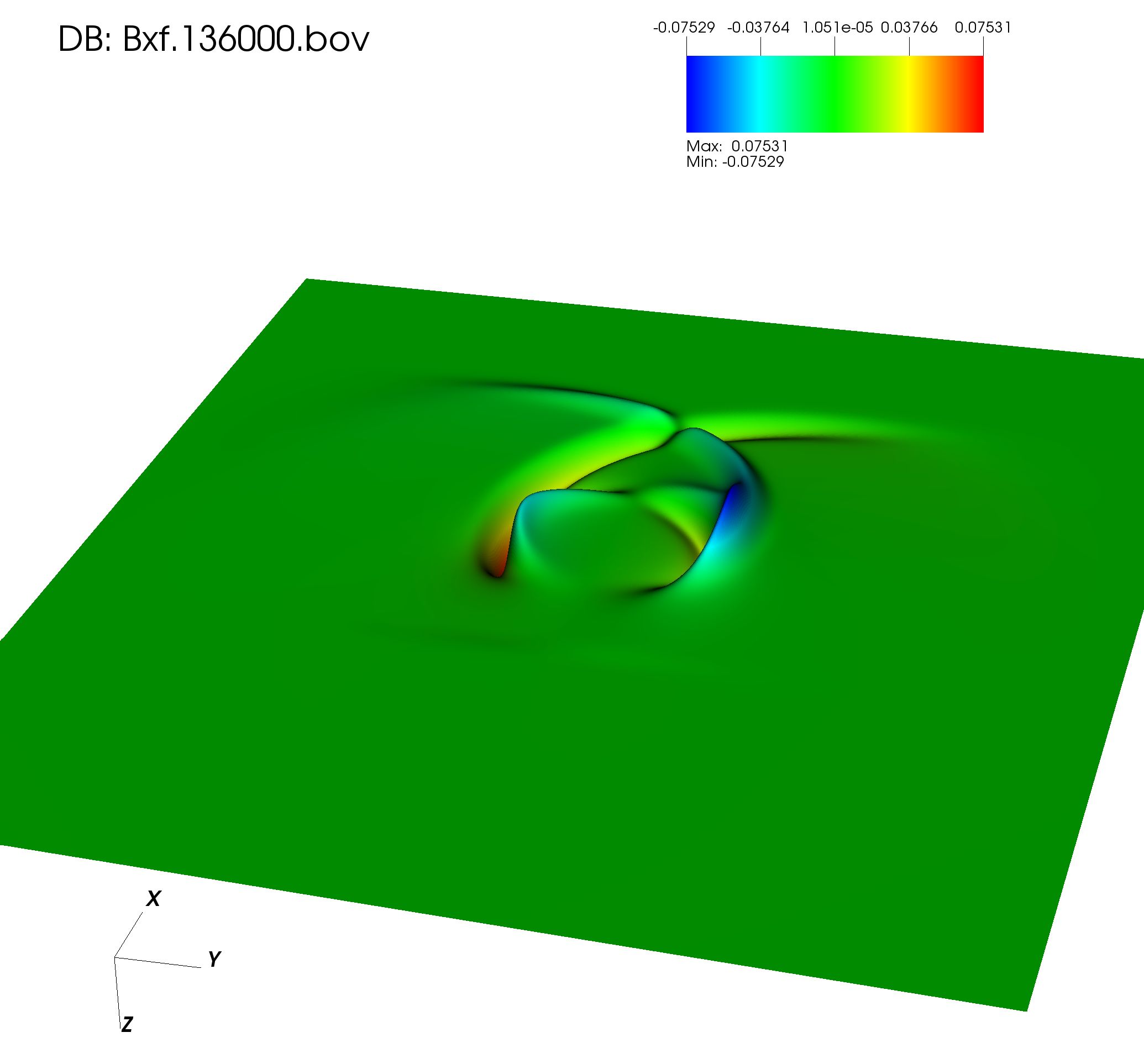}
(a) $B_x(x,y,t_0) > 0 $ at $t_0 = 36k$   \qquad ,  \qquad  \qquad  (b)  $B_x(x,y,t_0) < 0$ at $t_0 = 36k$
\caption{The spontaneously $B_x$ field at time $t = 36k$ that is generated by the QLA so that $\nabla \cdot \mathbf{B} = 0 $.  
This time corresponds to the $E_z$ field in Fig. 5, and $B_y$ field in FIg. 6.  The dipole structure of $B_x$ is clear on comparing
(a) and (b).  The perspective
(b) is obtained from (a) by rotating by $\pi$ about the line $y=L/2$.
}
\end{center}
\end{figure}
We find in our QLA simulations, that $max_{x,y} \left[ \nabla \cdot \mathbf{B} / |\mathbf{B}|  < 10^{-3} \right]$

\subsection{Scattering of 1D pulse with $E_y$ polarization}
We now turn to the 1D pulse with $E_y$ polarization, propagating in the $x-$direction toward the 2D tensor dielectric object, Fig. 1.  The other non-zero vacuum electromagnetic field
is $B_z(x,t)$.   On interacting with the tensor dielectric $\mathbf{n}(x,y)$, the scattered fields will develop a spatial dependence on $(x,y)$.  Thus $\nabla \cdot \mathbf{B} = 0$
exactly, and no new magnetic filed components need be generated,  
This is recognized by the QLA and so the only non-zero magnetic field throughout the run is $B_z(x,y,t)$.

In Fig. 8 we plot the $E_y$-field at time $t=18k$, the same time snapshot as for the case of $E_z$ polarization, Fig. 3.  The significant differences in the scattered field arise 
from the differences between the cylinder dielectric dependence of $n_y(x,y)$ and the cone $n_z(x,y)$.

 \begin{figure}[!h!p!b!t] \ 
\begin{center}
\includegraphics[width=3.2in]{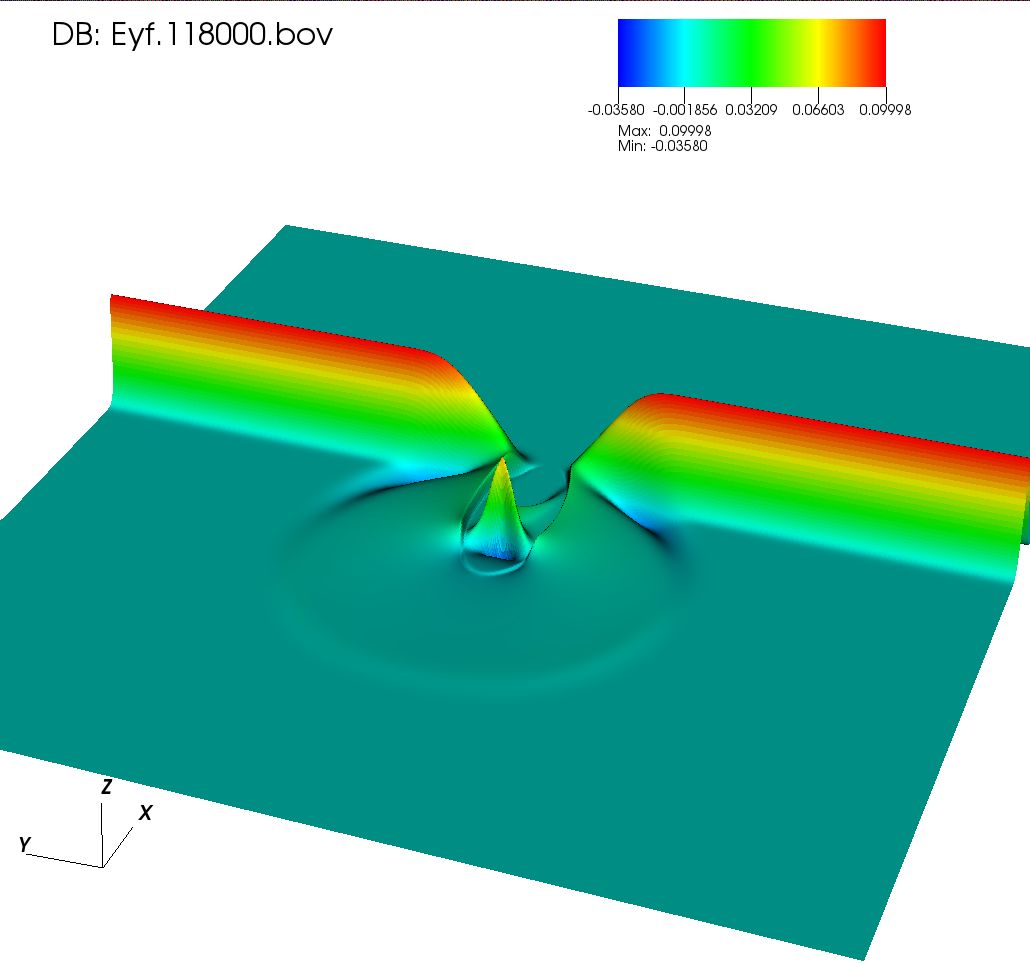}
\includegraphics[width=3.2in]{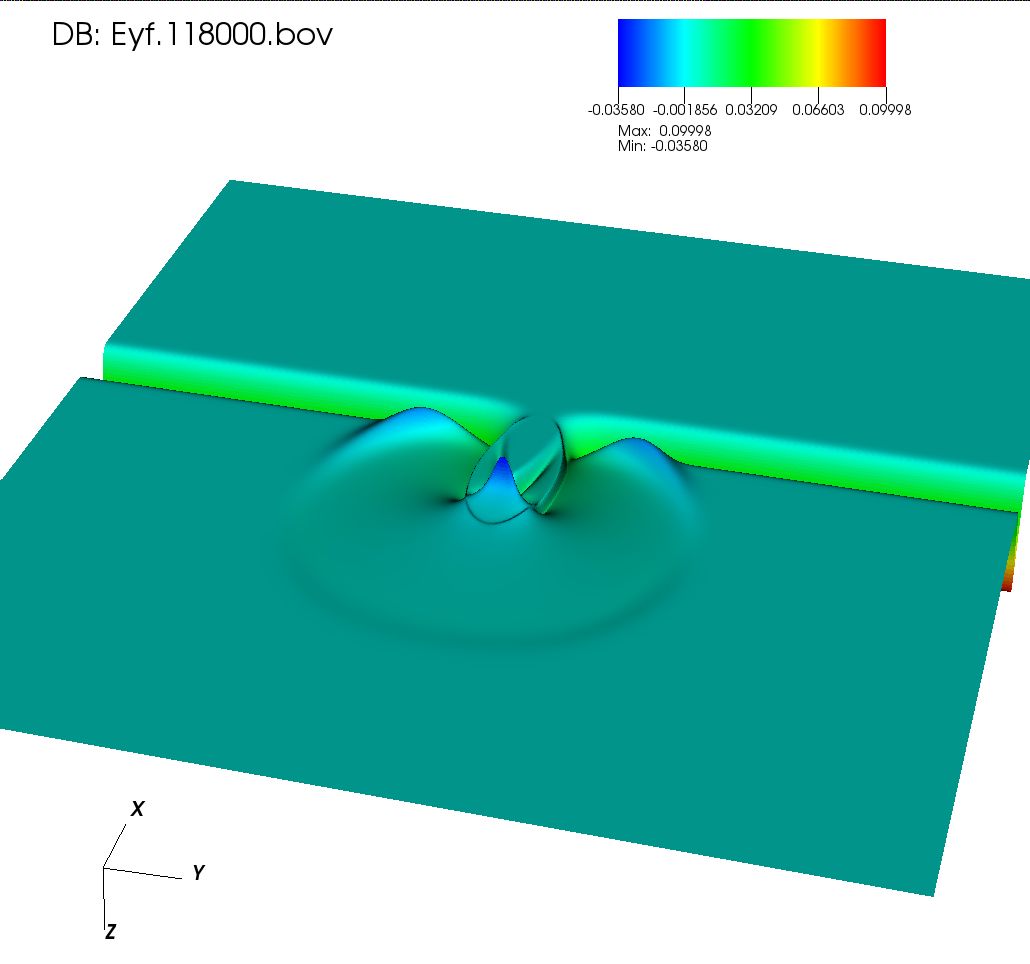}
(a) $E_y(x,y,t_0) > 0 $ at $t_0 = 18k$   \qquad ,  \qquad  \qquad  (b)  $E_y(x,y,t_0) < 0$ at $t_0 = 18k$
\caption{$E_y$ after interacting with the localized tensor dielectric.  Since the cylindrical $n_y$ dielectric has a sharper boundary layer than the conic $n_z$ dielectric, 
there is now a marked "reflected" wavefront propagating into the vacuum region together with the "transmitted" part of the pulse into the dielectric region itself.  
This "reflected" wavefront is absent when the major scattering is off the conic dielectric component, Fig. 3.  The perspective
(b) is obtained from (a) by rotating by $\pi$ about the line $y=L/2$.
}
\end{center}
\end{figure}
Also, what can be seen in Fig. 8 is the outward propagating circular-like wavefront which seems to be reminiscent of the reflected pulse in 1D scattering.  In particular, one 
sees elements of a $\pi$ phase change in this reflected wavefront.

The corresponding $E_y$ wavefronts at $t = 36k$ are shown in Fig. 9
 \begin{figure}[!h!p!b!t] \ 
\begin{center}
\includegraphics[width=3.2in]{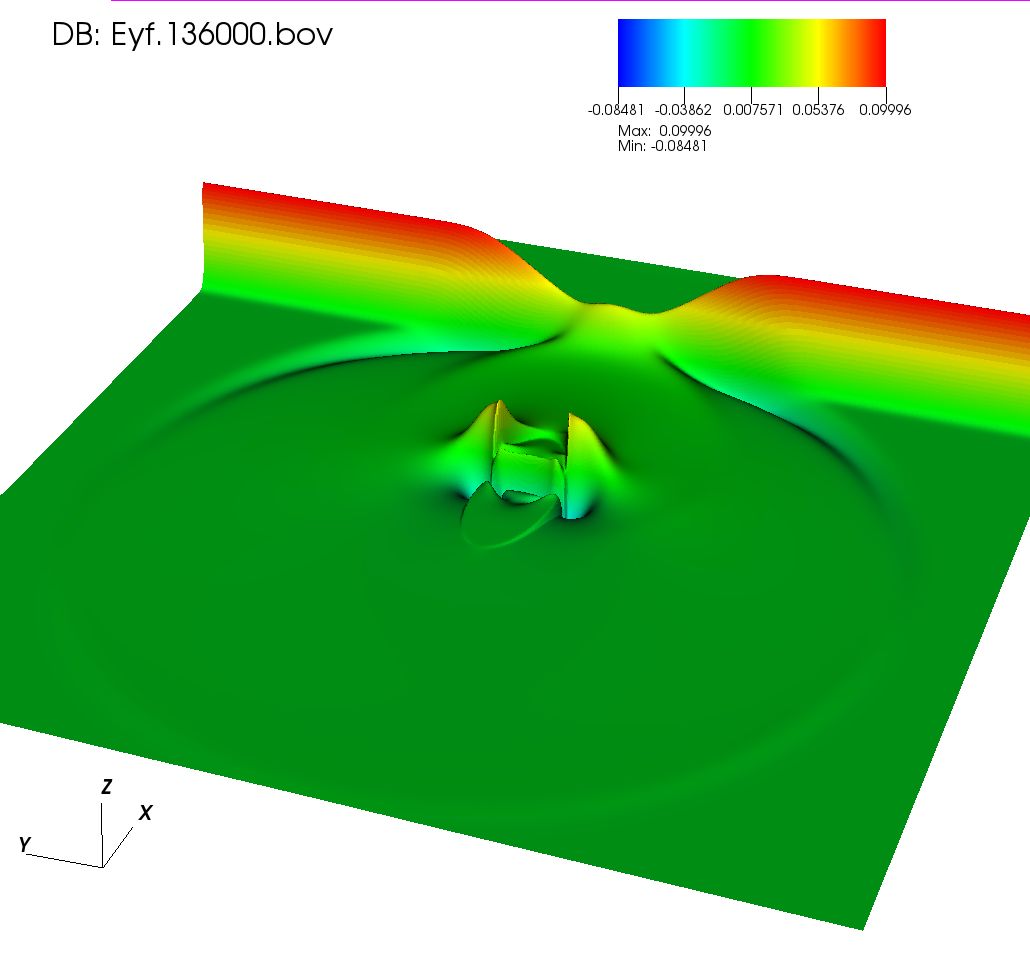}
\includegraphics[width=3.2in]{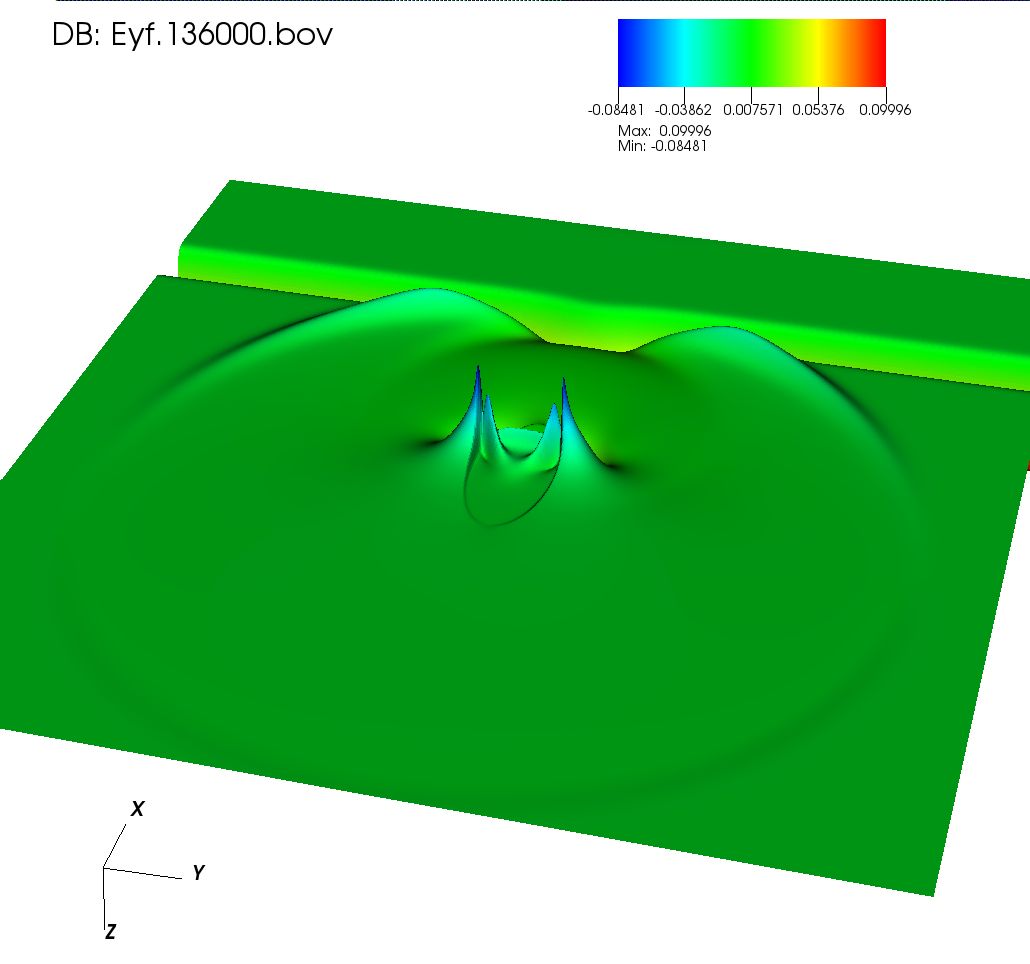}
(a) $E_y(x,y,t_2) > 0 $ at $t_2 = 36k$   \qquad ,  \qquad  \qquad  (b)  $E_y(x,y,t_2) < 0$ at $t_2 = 36k$
\caption{The $E_y$ wavefronts at a late stage of development, as the "reflected" pulse is about to reach the lattice boundaries.  The perspective
(b) is obtained from (a) by rotating by $\pi$ about the line $y=L/2$.
}
\end{center}
\end{figure}

The accompanying $B_z$ field of the initial 1D electromagnetic pulse is shown after its scattering from the tensor dielectric at times $t = 18k$, Fig 10, and at $t = 36k$, Fig. 11

 \begin{figure}[!h!p!b!t] \ 
\begin{center}
\includegraphics[width=3.2in]{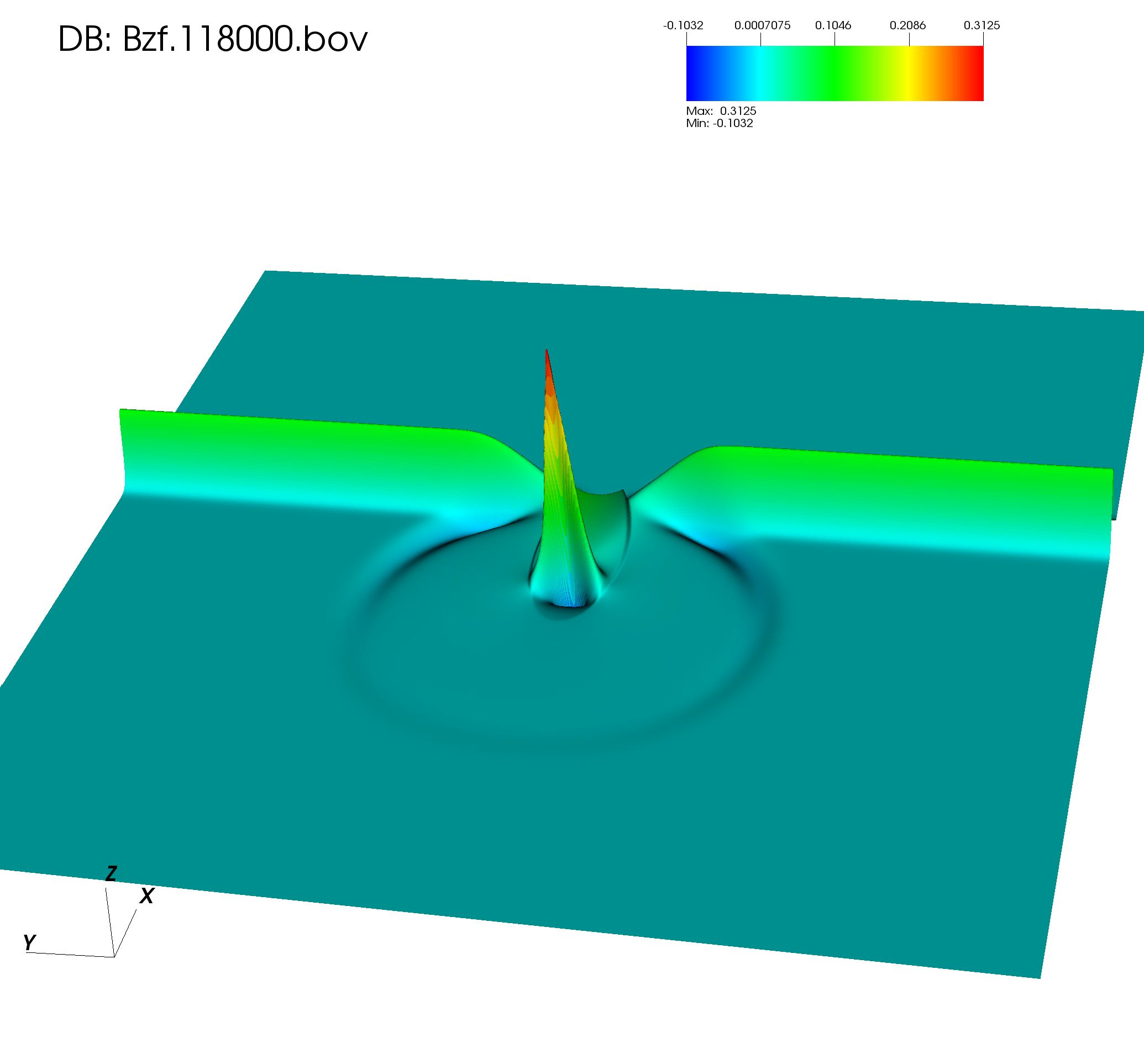}
\includegraphics[width=3.2in]{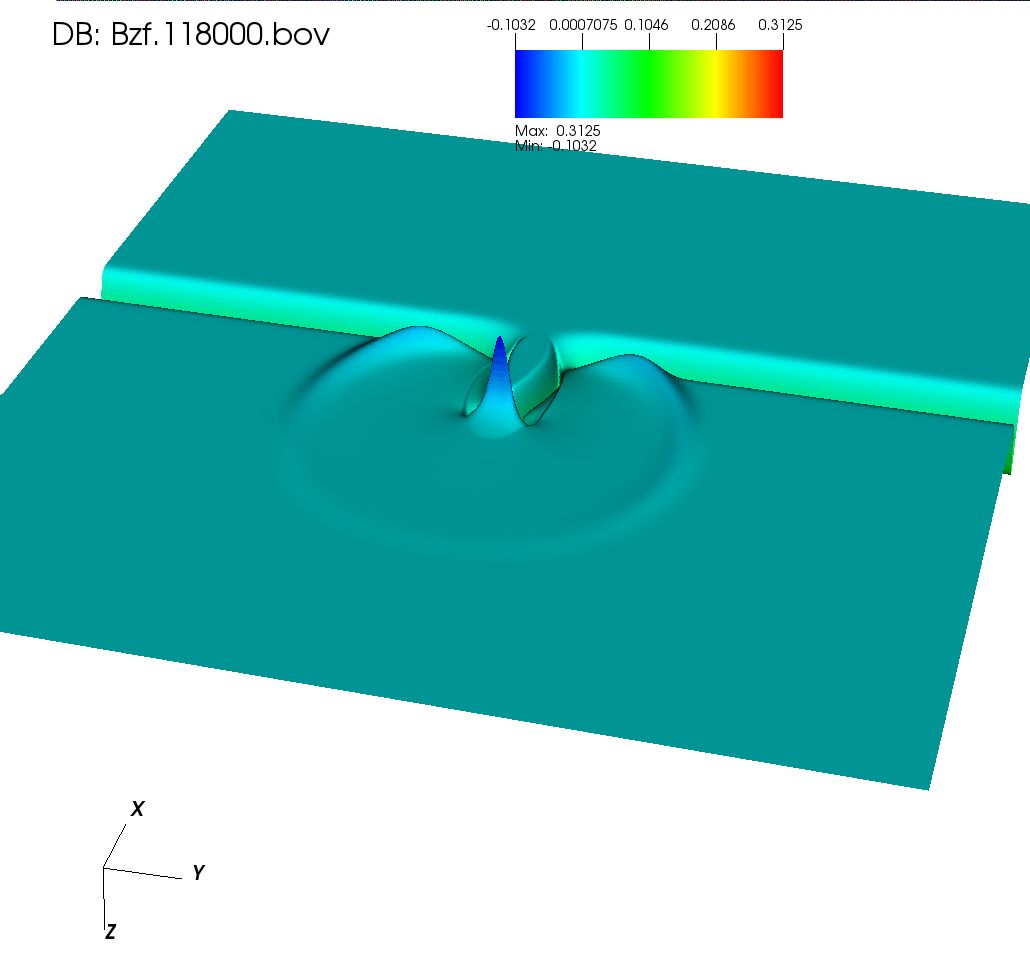}
(a) $B_z(x,y,t_1) > 0 $ at $t_1 = 18k$   \qquad ,  \qquad  \qquad  (b)  $B_z(x,y,t_1) < 0$ at $t_1 = 18k$
\caption{The $B_z$ wavefronts corresponding to the $E_y$ - field in Fig. 8.  The perspective
(b) is obtained from (a) by rotating by $\pi$ about the line $y=L/2$.
}
\end{center}
\end{figure}

 \begin{figure}[!h!p!b!t] \ 
\begin{center}
\includegraphics[width=3.2in]{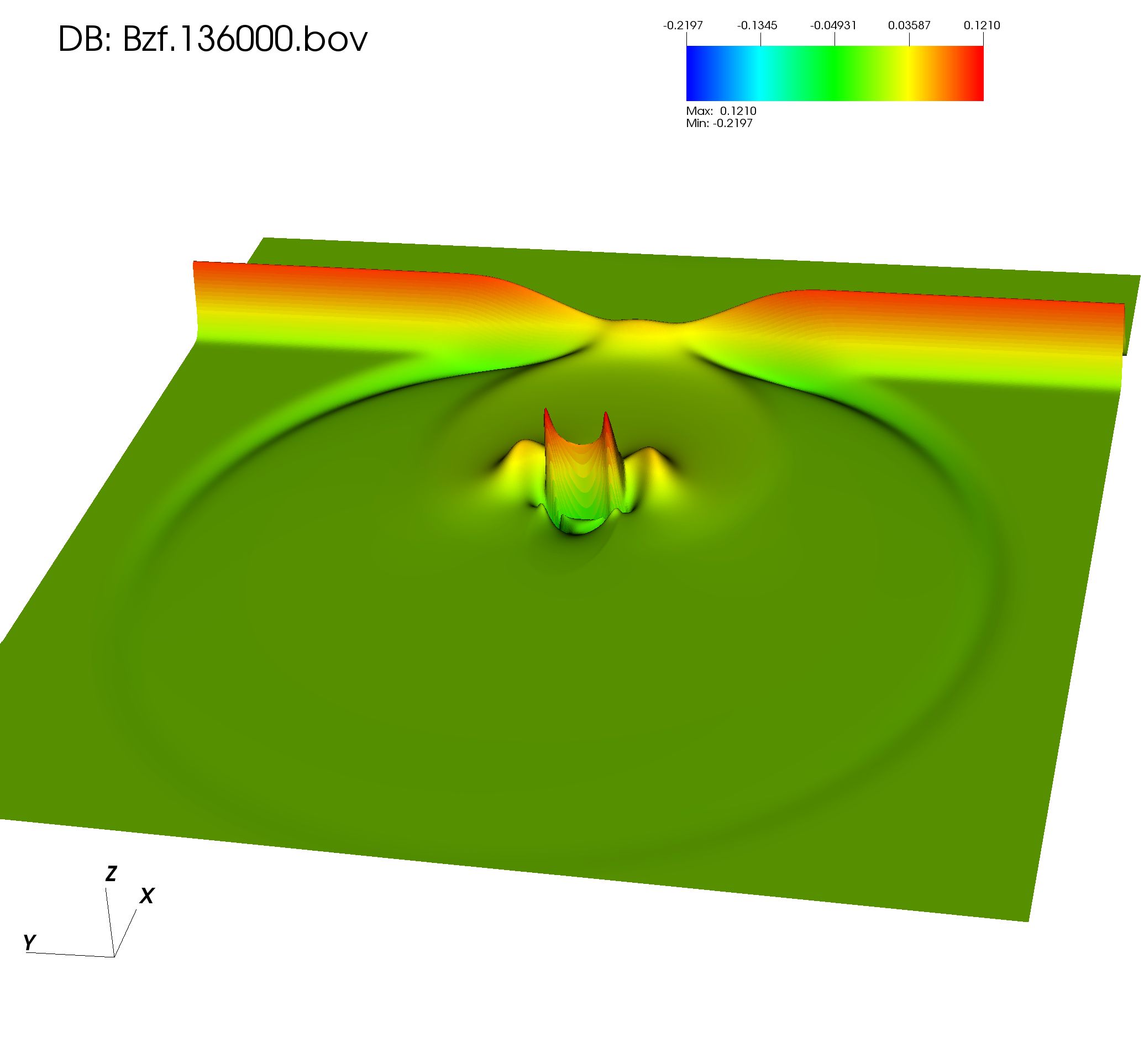}
\includegraphics[width=3.2in]{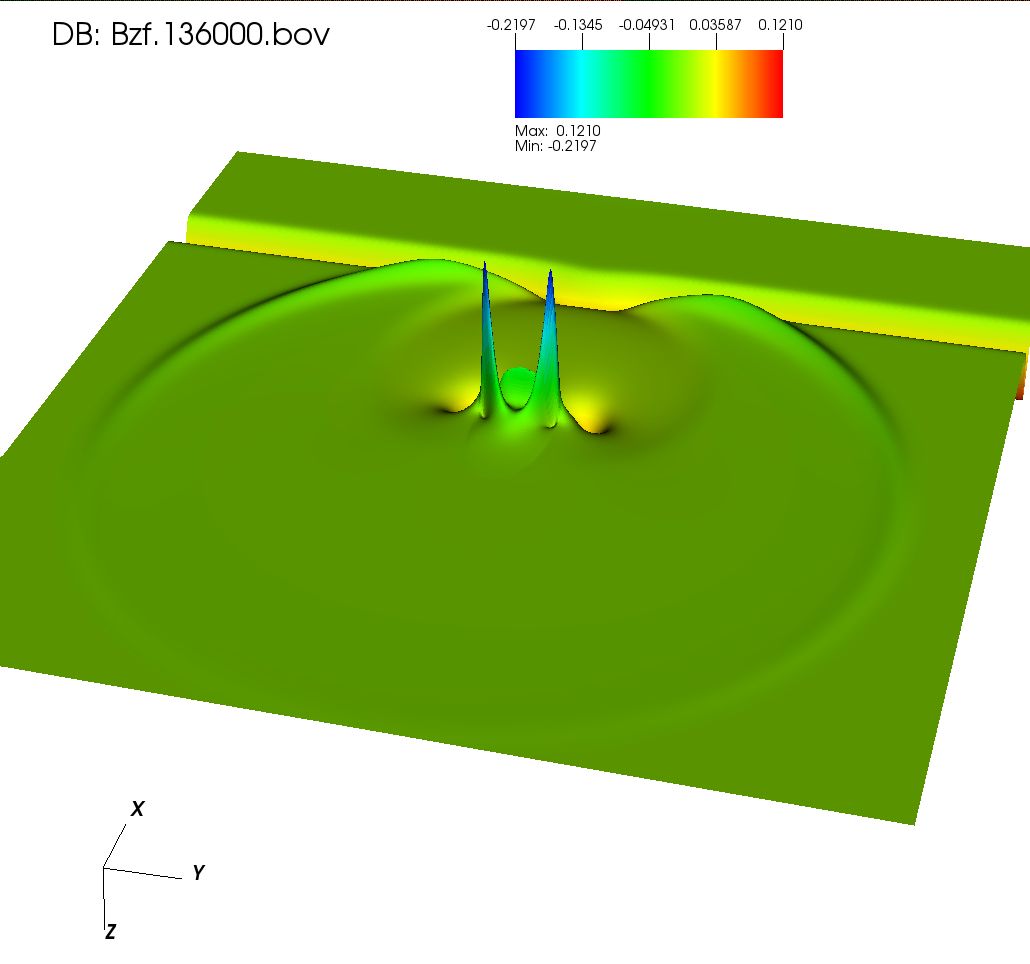}
(a) $B_z(x,y,t_2) > 0 $ at $t_2 = 36k$   \qquad ,  \qquad  \qquad  (b)  $B_z(x,y,t_2) < 0$ at $t_2 = 36k$
\caption{The $B_z$ wavefronts at a late stage of development, as the "reflected" pulse is about to reach the lattice boundaries.  The corresponding $E_y$ field 
is shown in Fig. 9.  The perspective
(b) is obtained from (a) by rotating by $\pi$ about the line $y=L/2$.
}
\end{center}
\end{figure}

Finally we consider the last of the Maxwell equations to be enforced:  $\nabla \cdot \mathbf{D} = 0$.  The QLA established a qubit basis for the curl-curl subset of Maxwell equations.  
For the initial polarization $E_z(x,t)$ and refractive indices $\mathbf{n=n}(x,y)$, the $\nabla \cdot \mathbf{D} = 0$ is automatically satisfied, while $\nabla \cdot \mathbf{B} = 0$ was spontaneously satisfied by the self-consistent generation of a $B_x$ field.  Now, if the initial polarization was $E_y(x,t)$, then $\nabla \cdot \mathbf{B} = 0$ is
automatically satisfied, while a spontaneously generated $E_x$ field is generated by the QLA  so that $\nabla \cdot \mathbf{D} = 0$ is satisfied.  In Fig. 12 we show the wavefronts of the $E_x$ field at time $t = 18k$ 
  \begin{figure}[!h!p!b!t] \ 
\begin{center}
\includegraphics[width=3.2in]{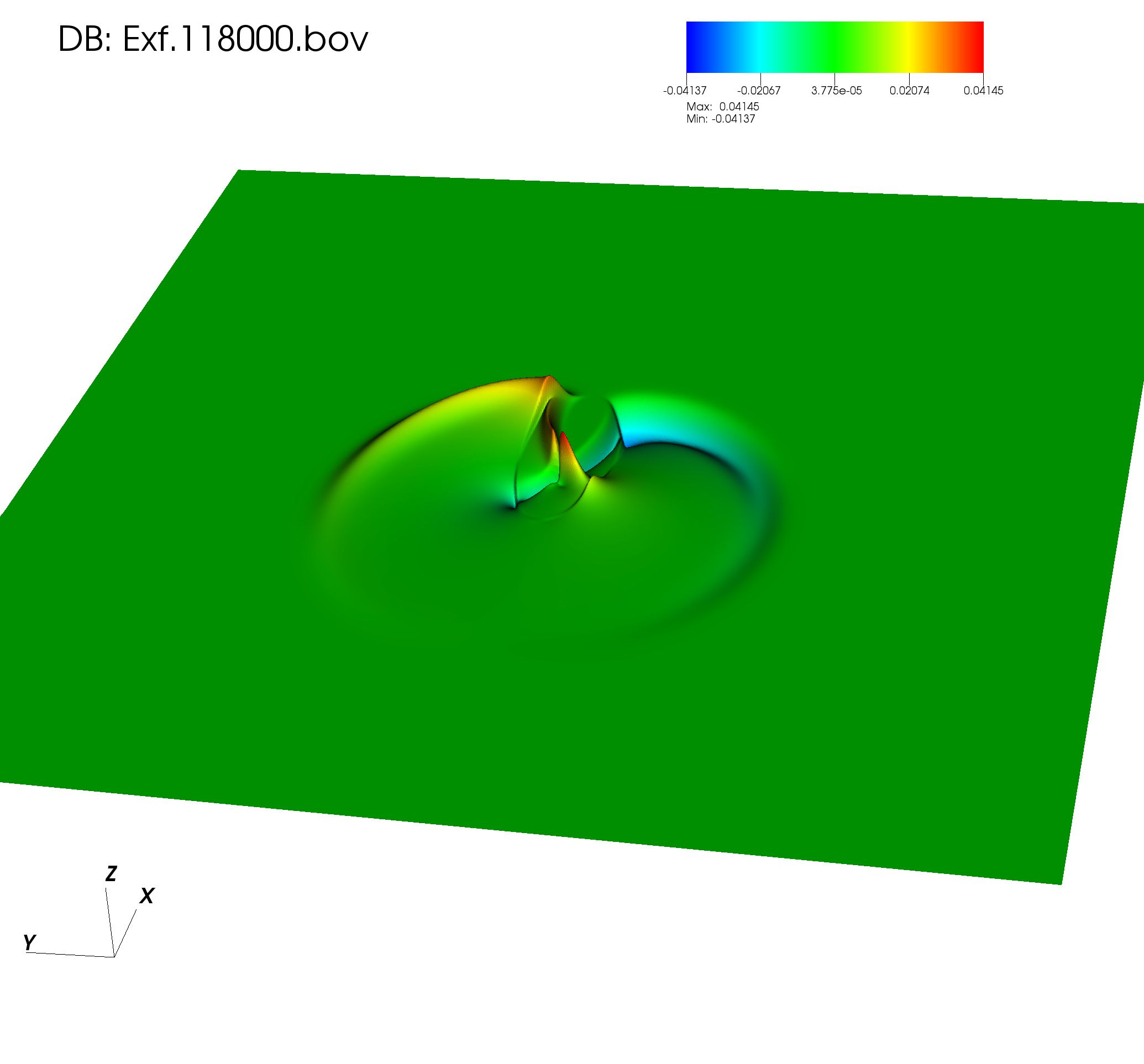}
\includegraphics[width=3.2in]{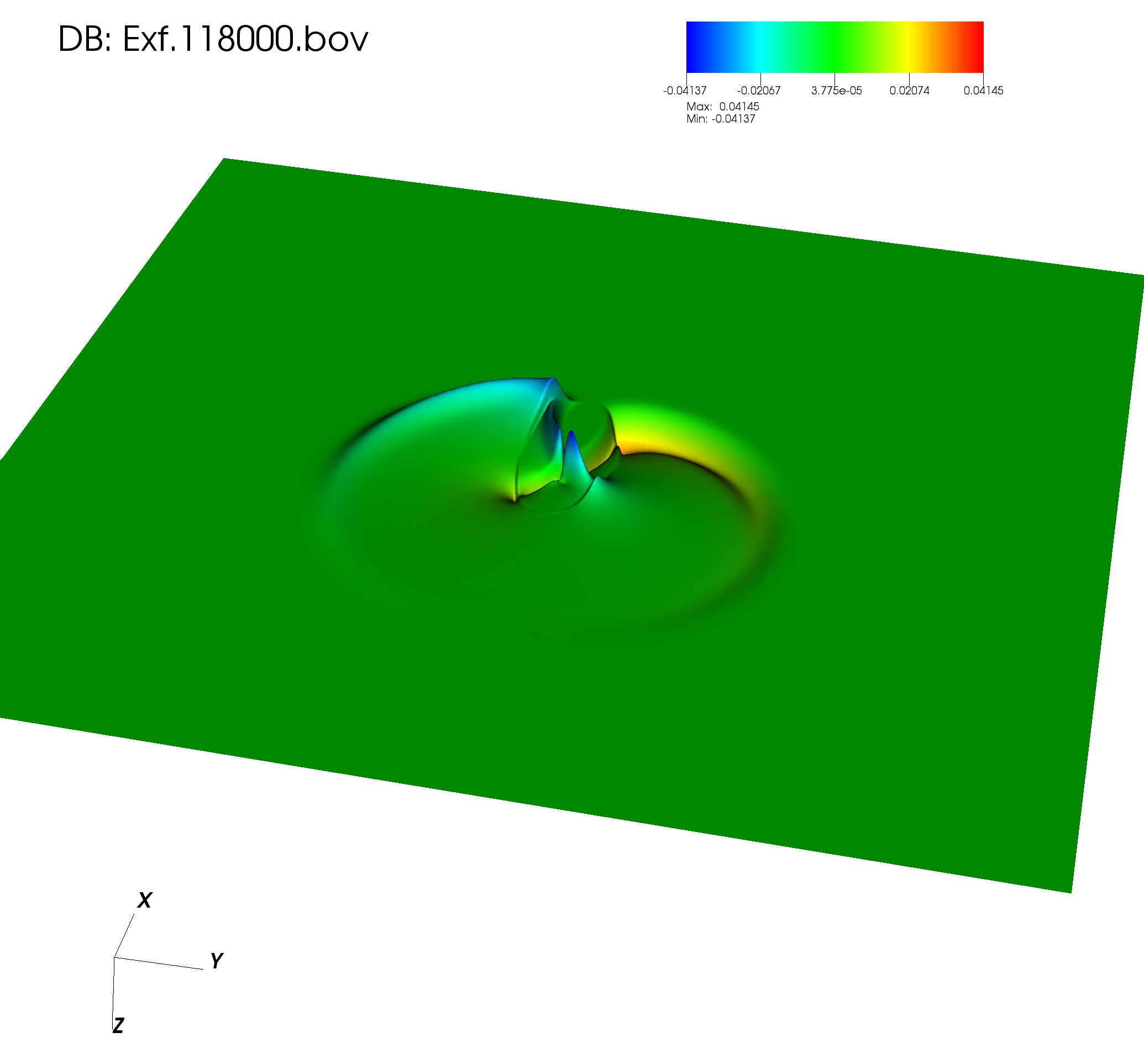}
(a) $E_x(x,y,t_1) > 0 $ at $t_1 = 18k$   \qquad ,  \qquad  \qquad  (b)  $E_x(x,y,t_1) < 0$ at $t_1 = 18k$
\caption{The $E_x$ wavefronts at time $18k$ spontaneously generated by QLA so as to (implicitly) satisfy the Maxwell equation $\nabla \cdot \mathbf{D} = 0$.
As the $E_x < 0$ plot perspective is generated from the $E_x > 0$ plot by rotating about the $y=L/2$ axis through an angle $\pi$, it is immediately seen
the the $E_x$ field strongly exhibits dipole structure.
}
\end{center}
\end{figure}
 
 \section{Summary}
 Determining a Dyson map, we have been able to develop a required basis from which the evolution equations for Maxwell equations can be unitary.
 In particular, we have shown that for inhomogeneous non-magnetic dielectric media, the field basis $(\mathbf{E,B})$ will not lead to a unitary representation.
 However, a particular Dyson map  shows that $(\mathbf{n . E, B})$, where $\mathbf{n}$ is a diagonal tensor dielectric, is a basis for a unitary representation.
 Other unitary representations can be immediately determined from this basis by unitary transformation, in particular the Riemann-Silberstein-Weber basis.
 
 Here we have concentrated on the basis $(\mathbf{n . E, B})$, primarily because the fields are real and so lead to quicker computations.  Our QLA directly
 encodes these fields into qubit representation.  A unitary set of interleaved collision-streaming operators are then applied to these qubits:  the unitary
 collision operators entangle the qubits, while the streaming operators move this entanglement throughout the lattice.  With our current set of unitary collision-streaming
 operators, we do not generate the effects of derivatives on the inhomogeneous medium.  These effects are included by the introduction of external potential
 operators - but at the expense of loosing the unitarity of the complete algorithm.
 
 In this paper we have performed QLA simulations on 2D scattering of a 1D electromagnetic pulse from a localized Hermitian tensor dielectric object.  Both polsrizations are considered with different field evolutions because of the anisotropic in the tensor dielectric.  The QLA we consider here are based on the two curl equations of Maxwell.
 Moreover the QLA is a perturbative representation, with small parameter $\delta$ representative of the spatial lattice width, with $QLA \rightarrow curl-curl-Maxwell$ as
 $\delta \rightarrow 0$.  It is not at all obvious that the QLA has the right structure to recover Maxwell equations - but only through symbolic manipulations (Mathematica) do we determine this Maxwell limit.  Hence it is of some interest to see how well QLA satisfies to two divergence equations of Maxwell that are not directly encoded in the
 QLA process.  We find spontaneous generation in the QLA so that $\nabla \cdot \mathbf{B} = 0$, $\nabla \cdot \mathbf{D} = 0$.
 
 Finally we comment on the conservation of energy $\mathcal{E}$:
 \begin{align*}
\mathcal{E}(t) = \frac{1}{L^2} \int_0^L \int_0^L dx dy \left[ n_x^2 E_x^2 + n_y^2 E_y^2 + n_z^2 E_z^2 + \mathbf{B}^2 \right] 
\end{align*}
 In QLA, $\mathcal{E} = \mathcal{E}(t,\delta)$.  Under appropriate scaling of the QLA operator angles, one recovers perturbatively the curl-curl Maxwell equations as 
 $\delta \rightarrow 0$.  Moreover, we find that $\mathcal{E}_{QLA} \rightarrow  const.$ as $\delta \rightarrow 0$.  The QLA simulations presented here were run on a lattice grid of $8192^2$, with $\delta = 0.1$.   
 
 The next step is to determine a fully unitary QLA for the Maxwell equations in anisotropic media.  The conservation of energy would be automatically satisfied as the norm 
 of the qubit basis.
 This unitary would then permit the QLA to be immediately encodable on a 
 quantum computer.  In the meantime, while we await error-correcting qubits and long decoherence time quantum computes, our current QLA's are ideally parallelized
 on classical supercomputers without core saturation effects.
 
 \section{Acknowledgments}
This research was partially supported by Department of Energy grants DE-SC0021647, DE-FG0291ER-54109, DE-SC0021651, DE-SC0021857, and DE-SC0021653. This work has been carried out partially within the framework of the EUROfusion Consortium. E.K has received funding from the Euratom research and training program WPEDU under grant agreement no. 101052200 as well as from the National Program for Controlled Thermonuclear Fusion, Hellenic Republic. K.H is
supported by the National Program for Controlled Thermonuclear Fusion, Hellenic Republic. The views and opinions expressed herein do not necessarily reflect those of the European Commission.

\section{References}
\qquad [1] VAHALA, G, VAHALA, L \& YEPEZ, J. 2003 Quantum lattice gas representation of some classical solitons. Phys. Lett A310, 187-196

[2] VAHALA, L, VAHALA, G \& YEPEZ, J. 2003 Lattice Boltzmann and quantum lattice gas representations of one-dimensional magnetohydrodynamic turbulence. Phys. Lett A306, 227-234.

[3] VAHALA, G, VAHALA, L \& YEPEZ, J. 2004. Inelastic vector soliton collisions: a latticebased quantum representation. Phil. Trans: Mathematical, Physical and Engineering Sciences, The Royal Society, 362, 1677-1690 [4] VAHALA, G, VAHALA, L \& YEPEZ, J. 2005 Quantum lattice representations for vector solitons in external potentials. Physica A362, 215-221.

[5] YEPEZ, J. 2002 An efficient and accurate quantum algorithm for the Dirac equation. arXiv: 0210093.

[6] YEPEZ, J. 2005 Relativistic Path Integral as a Lattice-Based Quantum Algorithm. Quant. Info. Proc. 4, 471-509.

[7] YEPEZ, J, VAHALA, G \& VAHALA, L. 2009a Vortex-antivortex pair in a Bose-Einstein condensate, Quantum lattice gas model of theory in the mean-field approximation. Euro. Phys. J. Special Topics 171, 9-14

[8] YEPEZ, J, VAHALA, G, VAHALA, L \& SOE, M. 2009b Superfluid turbulence from quantum Kelvin wave to classical Kolmogorov cascades. Phys. Rev. Lett. 103, 084501.

[9] VAHALA, G, YEPEZ, J, VAHALA, L, SOE, M, ZHANG, B, \& ZIEGELER, S. 2011 Poincare recurrence and spectral cascades in three-dimensional quantum turbulence. Phys. Rev. E84, 046713

[10] VAHALA, G, YEPEZ, J, VAHALA, L \&SOE, M, 2012 Unitary qubit lattice simulations of complex vortex structures. Comput. Sci. Discovery 5, 014013

[11] VAHALA, G, ZHANG, B, YEPEZ, J, VAHALA. L \& SOE, M. 2012 Unitary Qubit Lattice Gas Representation of 2D and 3D Quantum Turbulence. Chpt. 11 (pp. 239 - 272), in Advanced Fluid Dynamics, ed. H. W. Oh, (InTech Publishers, Croatia)

[12] YEPEZ, J. 2016 Quantum lattice gas algorithmic representation of gauge field theory. SPIE 9996, paper 9996-2

[13] OGANESOV, A, VAHALA, G, VAHALA, L, YEPEZ, J \& SOE, M. 2016a. Benchmarking the Dirac-generated unitary lattice qubit collision-stream algorithm for 1D vector Manakov soliton collisions. Computers Math. with Applic. 72, 386

[14] OGANESOV, A, FLINT, C, VAHALA, G, VAHALA, L, YEPEZ, J \& SOE, M 2016b Imaginary time integration method using a quantum lattice gas approach. Rad Effects Defects Solids $171,96-102$

[15] OGANESOV, A, VAHALA, G, VAHALA, L \& SOE, M. 2018. Effects of Fourier Transform on the streaming in quantum lattice gas algorithms. Rad. Eff. Def. Solids, 173, 169-174

[16] VAHALA, G., SOE, M., VAHALA, L., \& RAM, A. K., 2021 One- and Two-Dimensional quantum lattice algorithms for Maxwell equations in inhomogeneous scalar dielectric media I : theory. Rad. Eff. Def. Solids 176, 49-63.

[17] VAHALA, G., SOE, M., VAHALA, L., \& RAM, A. K., 2021 One- and Two-Dimensional quantum lattice algorithms for Maxwell equations in inhomogeneous scalar dielectric media II : Simulations. Rad. Eff. Def. Solids 176, 64-72.

[18] VAHALA, G, VAHALA, L, SOE, M \& RAM, A, K. 2020. Unitary Quantum Lattice Simulations for Maxwell Equations in Vacuum and in Dielectric Media, J. Plasma Phys 86, 905860518

[19] VAHALA, L, SOE, M, VAHALA, G \& YEPEZ, J. 2019a. Unitary qubit lattice algorithms for spin-1 Bose-Einstein condensates. Rad Eff. Def. Solids 174, 46-55

[20] VAHALA, L, VAHALA, G, SOE, M, RAM, A \& YEPEZ, J. 2019b. Unitary qubit lattice algorithm for three-dimensional vortex solitons in hyperbolic self-defocusing media. Commun Nonlinear Sci Numer Simulat 75, 152-159

[21] RAM, A. K., VAHALA, G., VAHALA, L. \& SOE, M 2021 Reflection and transmission of electromagnetic pulses at a planar dielectric interface - theory and quantum lattice simulations AIP Advances 11, 105116 (1-12). 

[22] MERMIN, N. D., 2007 Quantum computer science, Cambridge University Press, Cambridge 

[23] KOUKOUTSIS, E., HIZANIDIS, K., RAM, A. K., \& VAHALA, G. 2022. Dyson Maps and Unitary Evolution for Maxwell Equations in Tensor Dielectric Media. arXiv:2209.08523

 \end{document}